\documentclass[twocolumn,superscriptaddress,amsfont,amssymb,amsmath, showpacs,balancelastpage, nofootinbib]{revtex4-1}
\usepackage{graphicx,longtable,mathrsfs,color,array}
\usepackage[unicode=true,pdfusetitle,
bookmarks=true,bookmarksnumbered=true,bookmarksopen=true,bookmarksopenlevel=1,
breaklinks=false,pdfborder={0 0 0},backref=false,colorlinks=true]{hyperref}
\hypersetup{citecolor=blue,filecolor=blue,linkcolor=blue,urlcolor=blue}
\usepackage[usenames,dvipsnames]{xcolor}
\usepackage{amssymb,amsmath,mathtools,mathrsfs,enumitem}
\usepackage{epsfig,subfigure,placeins,float}
\usepackage{booktabs,longtable,multirow}
\usepackage{exscale,relsize}
\usepackage[normalem]{ulem}
\usepackage[T1]{fontenc}
\usepackage[utf8]{inputenc}
\usepackage{enumerate}
\usepackage{times, mathptmx}
\usepackage{tikz}
\usepackage{aas_macros}
\usetikzlibrary{arrows,positioning,decorations.markings,decorations.pathmorphing,calc}

\newcommand{\be}{\begin{equation}}
\newcommand{\ee}{\end{equation}}
\newcommand{\cgw}{c_{\rm GW}}
\newcommand{\MPl}{M_{\rm Pl}}
\newcommand{\Oo}{{{\cal O}(1)}}
\newcommand{\aH}{\alpha_H}
\newcommand{\aB}{\alpha_B}
\newcommand{\aM}{\alpha_M}
\newcommand{\aT}{\alpha_T}
\newcommand{\aK}{\alpha_K}
\newcommand{\aI}{\alpha_i}
\newcommand{\aQS}{\alpha_{\rm QS}}
\newcommand{\km}{{\rm km}}
\newcommand{\comment}[1]{}

\newcolumntype{C}[1]{>{\centering\let\newline\\\arraybackslash\hspace{0pt}}m{#1}}

\definecolor{hyperref}{RGB}{026,028,087}
\def\gsim{ \lower .75ex \hbox{$\sim$} \llap{\raise .27ex \hbox{$>$}} }
\def\lsim{ \lower .75ex \hbox{$\sim$} \llap{\raise .27ex \hbox{$<$}} }
\newcommand{\dof}{{\it dof} }

\def\nn{\nonumber}
\def\ni{\noindent}

\newcommand*{\mathcolor}{}
\def\mathcolor#1#{\mathcoloraux{#1}}
\newcommand*{\mathcoloraux}[3]{%
\protect\leavevmode
\begingroup
\color#1{#2}#3%
\endgroup
}
\newlength{\stheight}
\newcommand\textst[1][fu-grey]{
\ifmmode\setlength{\stheight}{+1.0ex}
\else\setlength{\stheight}{+0.5ex}
\fi
\bgroup\markoverwith{\textcolor{#1}{\rule[\the\stheight]{2pt}{1.0pt}}}\ULon
} % strikethrough text command
\newcommand{\textins}[2][fu-grey]{
\ifmmode\mathcolor{#1}{#2}
\else\textcolor{#1}{#2}\@\,
\fi
}
\graphicspath{{./Plots/}}
\allowdisplaybreaks

\tikzstyle{vecArrow} = [thick, decoration={markings,mark=at position
1 with {\arrow[semithick]{open triangle 60}}},
double distance=1.4pt, shorten >= 5.5pt,
preaction = {decorate},
postaction = {draw,line width=1.4pt, white,shorten >= 4.5pt}]

\begin{document}
\title{Cosmological constraints on dark energy in light of gravitational wave bounds}

\author{Johannes Noller}
\affiliation{Institute for Theoretical Studies, ETH Z\"urich, Clausiusstrasse 47, 8092 Z\"urich, Switzerland}
\affiliation{Institute for Particle Physics and Astrophysics, ETH Z\"urich, 8093 Z\"urich, Switzerland}

\begin{abstract}
Gravitational wave (GW) constraints have recently been used to significantly restrict models of dark energy and modified gravity. New bounds arising from GW decay and GW-induced dark energy instabilities are particularly powerful in this context, complementing bounds from the observed speed of GWs.  
We discuss the associated linear cosmology for Horndeski gravity models surviving these combined bounds and compute the corresponding cosmological parameter constraints, using CMB, redshift space distortion, matter power spectrum and BAO measurements from the Planck, SDSS/BOSS and 6dF surveys.
The surviving theories are strongly constrained, tightening previous bounds on cosmological deviations from $\Lambda{}$CDM by over an order of magnitude. 
We also comment on general cosmological stability constraints and the nature of screening for the surviving theories, pointing out that a raised strong coupling scale can ensure compatibility with gravitational wave constraints, while maintaining a functional Vainshtein screening mechanism on solar system scales. Finally, we discuss the quasi-static limit as well as (constraints on) related observables for near-future surveys. 
\end{abstract}

\date{\today}
\maketitle

\section{Introduction} \label{sec-intro}

Constraints derived from considering the interplay between gravitational waves (GWs) and other light gravitational degrees of freedom (dofs), potentially related to dark energy, have recently been argued to strongly restrict the latter. These constraints include measurements of the speed of GWs \cite{PhysRevLett.119.161101,2041-8205-848-2-L14,2041-8205-848-2-L15,2041-8205-848-2-L13,2041-8205-848-2-L12} (for a discussion of bounds derived from this measurement see \cite{Baker:2017hug,Creminelli:2017sry,Sakstein:2017xjx,Ezquiaga:2017ekz}), bounds on the decay of GWs into dark energy \cite{Creminelli:2018xsv,Creminelli:2019nok} and the requirement of the absence of GW-induced dark energy instabilities \cite{Creminelli:2019kjy}. Here we discuss what these constraints imply for cosmological parameter constraints on theories of dark energy (and/or modified gravity).
\\

\ni{\bf Horndeski gravity}: Since GR is the single consistent theory of a massless spin-2 field, testing for (potentially dark energy-related) deviations away from it amounts to probing the presence of new gravitational degrees of freedom. Scalar-tensor (ST) theories are a minimal deviation from GR in this sense, as they only introduce a single additional degree of freedom. Accordingly, Horndeski gravity \cite{Horndeski:1974wa,Deffayet:2011gz},\footnote{For the equivalence between the formulations of \cite{Horndeski:1974wa} and \cite{Deffayet:2011gz}, see \cite{Kobayashi:2011nu}.} the most general Lorentz-invariant ST action that gives rise to second order equations of motion, has recently been the main workhorse in testing for deviations from GR. 
It is described by the following action
\begin{eqnarray}\label{Horndeski_action}
S_H=\int \mathrm{d}^4x \sqrt{-g}\left\{\sum_{i=2}^5{\cal L}_i[\phi,g_{\mu\nu}]\right\},
\end{eqnarray}
where we write the scalar-tensor Lagrangians ${\cal L}_i$ (for a scalar $\phi$ and a massless tensor $g_{\mu\nu}$) as
\begin{align}
{\cal L}_{2} & = \Lambda_2^4 \, G_2~, \quad\quad\quad\quad {\cal L}_{3} = \frac{\Lambda_2^4}{\Lambda_3^3} G_{3}\cdot[\Phi]\,, \nn \\
{\cal L}_{4}  & = \frac{\Lambda_2^8}{\Lambda_3^6} G_{4} R + \frac{\Lambda_2^4}{\Lambda_3^6} ~ G_{4,X} \left( [\Phi]^2-[\Phi^2] \right)\,,   \label{Horndeski_lagrangians} \\
{\cal L}_{5} & = \frac{\Lambda_2^{8}}{\Lambda_3^9} G_{5}G_{\mu\nu}\Phi^{\mu\nu}-\frac{1}{6} \frac{\Lambda_2^4}{\Lambda_3^9} G_{5,X}([\Phi]^3 -3[\Phi][\Phi^2]+2[\Phi^3]), \nn
\end{align}
where $X = -\tfrac{1}{2}\nabla_\mu \phi \nabla^\mu \phi/\Lambda_2^4$ is the scalar kinetic term, $\Phi^{\mu}_{\;\; \nu} \equiv  \nabla^\mu \nabla_\nu\phi$, the $G_i$ are dimensionless functions of $\phi/\MPl$ and $X$, and $G_{i,\phi}$ and $G_{i,X}$ denote the partial derivatives of the $G_i$ (with respect to these dimensionless arguments). Square brackets denote the trace, e.g. $ [\Phi^2] \equiv \nabla^\mu \nabla_\nu \phi \nabla^\nu \nabla_\mu\phi$ and we have three mass scales: $\MPl,  \Lambda_2$ and $\Lambda$. In cosmology they are conventionally taken to satisfy $\Lambda_2 = \MPl H_0$ and $\Lambda_3 = \MPl H_0^2$, which ensures that all interactions can give $\Oo$ contributions to the background evolution.
\\

\ni{\bf Linear cosmology}:
Since we are focusing on large scale observables, we are particularly interested in linearised perturbations around a cosmological FRW background. 
For the general Horndeski theory \eqref{Horndeski_lagrangians}, the freedom in the dynamics of such perturbations is controlled by just four functions $\aI$ of time. More specifically, these $\aI$ are given by \cite{Bellini:2014fua}
\begin{align}
M^2 &= 2\left(G_4-2XG_{4,X}+XG_{5,\phi}-\frac{{\dot \phi}H}{H_0^2} XG_{5,X}\right) , \nonumber \\
M^2\aB &= - 2\frac{\dot{\phi}}{H}\left(XG_{3,X}+G_{4,\phi}+2XG_{4,\phi X}\right) \nonumber \\ & 
+8X\left(G_{4,X}+2XG_{4,XX}-G_{5,\phi}-XG_{5,\phi X}\right) \nonumber \\ & 
+2\frac{\dot{\phi}H}{H_0^2}X\left(3G_{5,X}+2XG_{5,XX}\right) \nonumber , \\ 
M^2\aT &= 2X\left[2G_{4,X}-2G_{5,\phi}-\left(\frac{\ddot{\phi}}{H_0^2}-\frac{\dot{\phi}H}{H_0^2}\right)G_{5,X}\right] \,,
\label{alphadef}
\end{align}
where we also define $HM^2\aM \equiv \tfrac{d}{dt}M^2$.\footnote{Note that there is a (conventional) sign difference for $G_3$ in the expressions for the $\alpha_i$ here compared to \cite{Bellini:2014fua}. This is due to a sign difference in our formulation of the Horndeski action \eqref{Horndeski_action} compared to the corresponding formulation of \cite{Bellini:2014fua}. Factor of $H_0$ differences are due to our dimensionless definition of the $G_i$ (as opposed to the dimensionful $G_i$ in \cite{Bellini:2014fua}).} Here $\aM$ is the ``running'' of the effective Planck mass $M_{\rm Pl}^{\rm eff} \equiv M \MPl$; $\aB$, the ``braiding'' that quantifies kinetic mixing between the metric and scalar perturbations; and $\aT$, the tensor speed excess, related to the sound speed of tensor perturbations via $c_{GW}^2 = 1 + \aT$. 
The kineticity $\aK$ is the fourth free function, but will be omitted here for now, since it does not affect constraints on other parameters at leading order at the level of linear perturbations  \cite{BelliniParam,Alonso:2016suf} -- we will get back to $\aK$ later. 
\\

\section{GW constraints} \label{sec-GWcon}

\ni{\bf Speed of gravity}: The deviation of the speed of gravitational waves (GWs) from that of light is quantified via $\aT \equiv \cgw^2 - 1$ (in natural units) and is explicitly given in \eqref{alphadef} above.
The near simultaneous detections of GW170817 and GRB 170817A \cite{PhysRevLett.119.161101,2041-8205-848-2-L14,2041-8205-848-2-L15,2041-8205-848-2-L13,2041-8205-848-2-L12} established that the speed of GWs and that of light differ by at most one part in $10^{15}$ at the energy/frequency scales probed by LIGO ($10 - 10^4$ Hz \cite{Martynov:2016fzi}). 
Since the frequencies observed for GW170817 are close to $\Lambda_3 \sim 10^2$ Hz, additional assumptions about the UV physics are necessary to apply the bound from this event to a candidate cosmological theory like \eqref{Horndeski_action} \cite{deRham:2018red}. This is because $\Lambda_3$ is also the naive cutoff for the theory, so new physics associated to an unknown UV completion is expected to enter at or below $\Lambda_3$ in order to unitarise the theory. Whether the bound from GW170817 can straightforwardly be mapped to cosmological scales then depends on the features of that UV completion. 
While these features are not known, here we will follow the approach of \cite{Baker:2017hug,Creminelli:2017sry,Sakstein:2017xjx,Ezquiaga:2017ekz} (see \cite{Amendola:2012ky,Amendola:2014wma,Deffayet:2010qz,Linder:2014fna,Raveri:2014eea,Saltas:2014dha,Lombriser:2015sxa,Lombriser:2016yzn,Jimenez:2015bwa,Bettoni:2016mij,Sawicki:2016klv} for earlier work related to $c_{\rm GW} = c$ constraints\footnote{Also see \cite{Copeland:2018yuh} for a (closed) loophole to the argument of \cite{Baker:2017hug,Creminelli:2017sry,Sakstein:2017xjx,Ezquiaga:2017ekz}.}) and assume a frequency-independent speed of GWs, but note that the speed of GWs will be probed experimentally at lower frequencies with future gravitational wave experiments (in particular, LISA probes the $10^{-4} - 1$ Hz range below $\Lambda_3$ \cite{Audley:2017drz}), so this assumption will be tested in the near-future. With this assumption the speed of GWs at cosmological scales then is required to satisfy
\begin{align} \label{ligo_bound}
|\aT| \lesssim 10^{-15}.
\end{align}
Imposing the luminal propagation of GWs in this way significantly restricts \eqref{Horndeski_lagrangians}. In order to comply with \eqref{ligo_bound} the terms in \eqref{alphadef} that contribute to an anomalous GW speed, i.e. $G_5$ and $G_{4,X}$, need to be suppressed at the $10^{-15}$ level, i.e. $\aT = 0$ to all intents and purposes as far as large scale cosmology is concerned. This then reduces the total Lagrangian for \eqref{Horndeski_lagrangians} to the following restricted theory \cite{Baker:2017hug,Creminelli:2017sry,Sakstein:2017xjx,Ezquiaga:2017ekz}
\begin{align} \label{Horndeski_simple}
{\cal L}_{2}=\Lambda_2^4 G_2(\phi,X) + \frac{\Lambda_2^4}{\Lambda_3^3}G_3(\phi,X)[\Phi] + \MPl^2 G_4(\phi) R.
\end{align}
This theory yields $\aT = 0$ by design, the effective Planck mass is now given by $M^2 = 2G_4$, while $\aM$ and $\aB$ satisfy 
\begin{align}
HM^2 \aM &= 2 \dot\phi G_{4,\phi} , &H M^2\aB &= - 2\dot{\phi}\left(XG_{3,X}+ G_{4,\phi}\right),
\label{alphadef2}
\end{align}
so $\aM$ and $\aB$ are still independent free functions for this subset of Horndeski theories.
\\

\begin{figure*}[t]
\includegraphics[width=.49\linewidth]{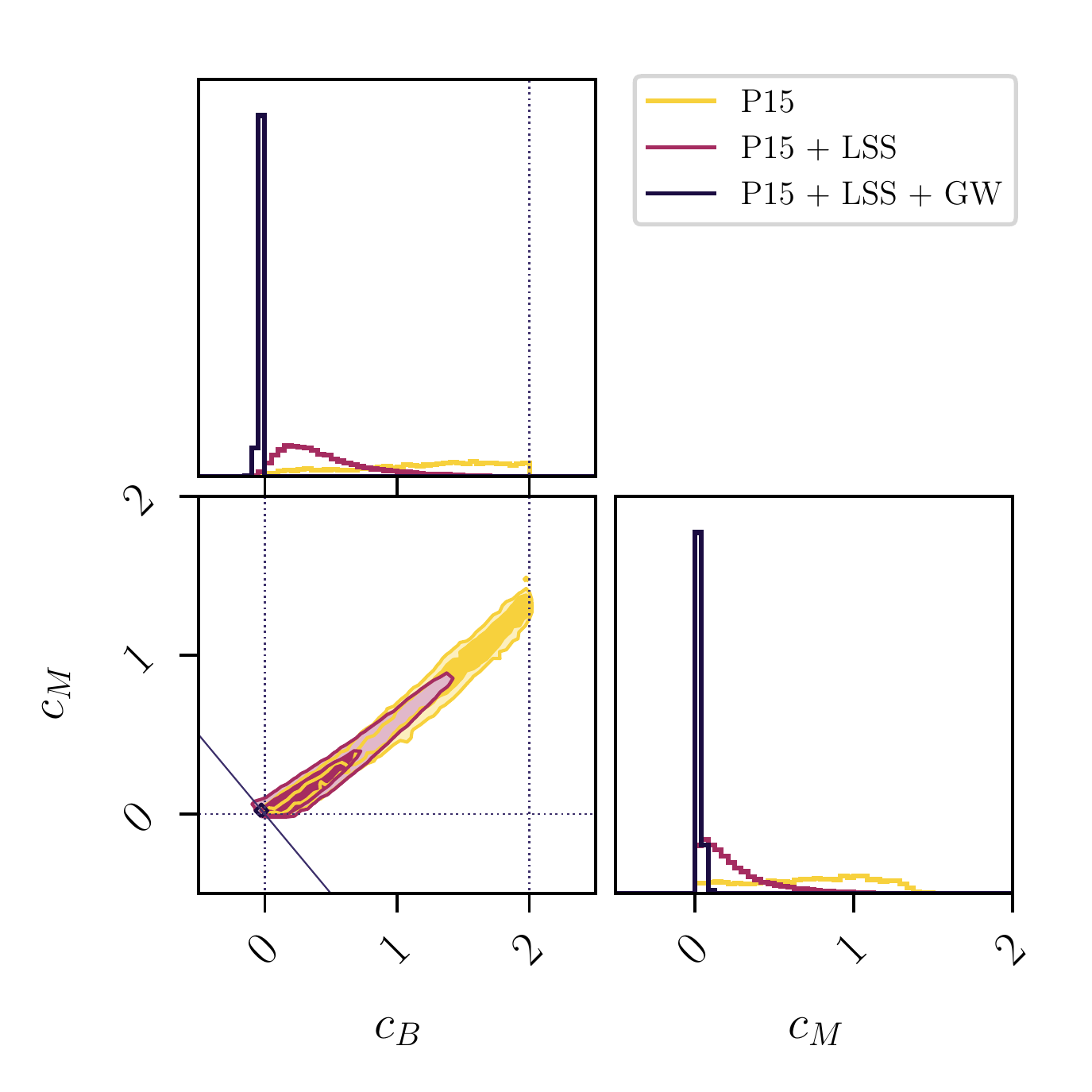}
\includegraphics[width=.49\linewidth]{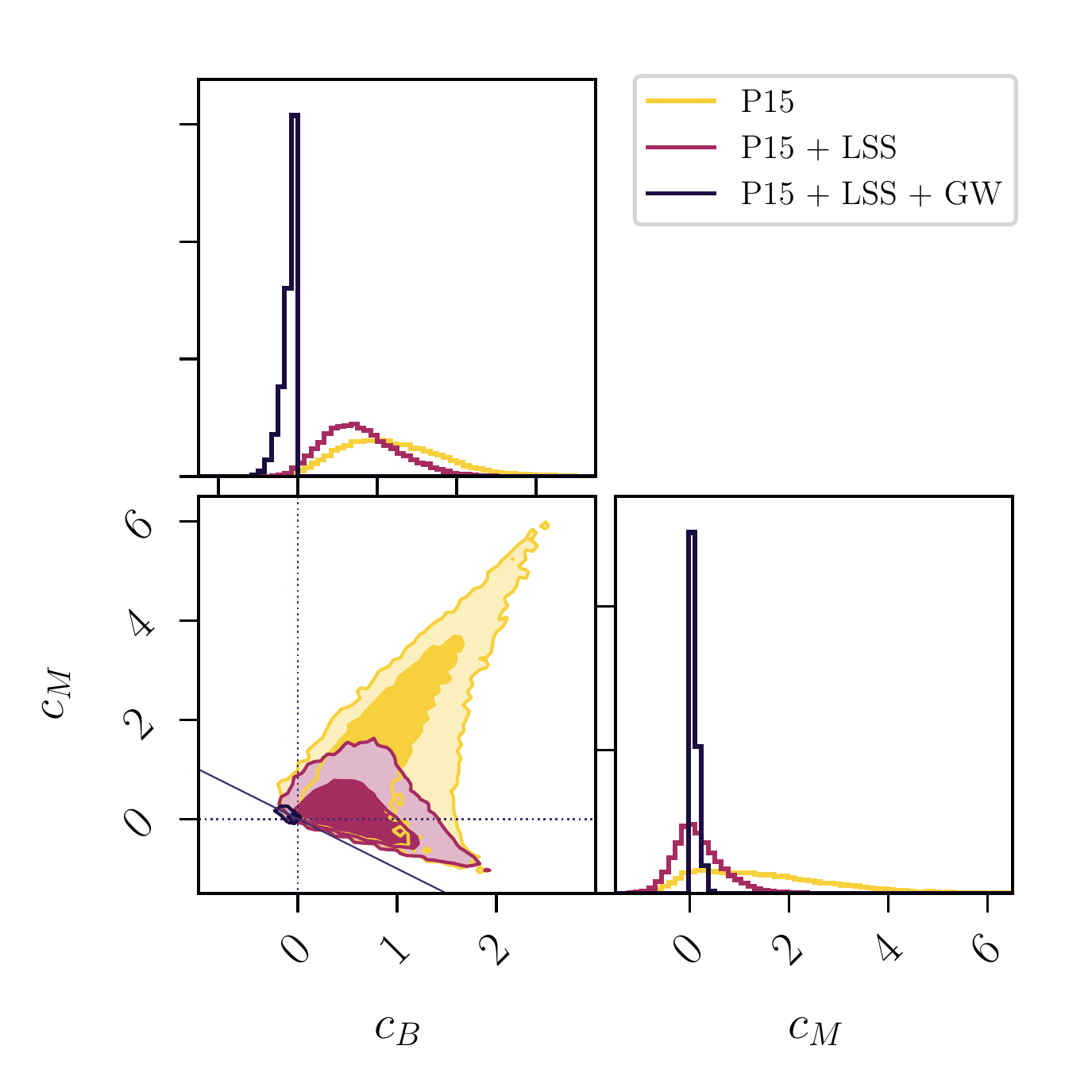}
\caption{
Cosmological parameter constraints for the dark energy $c_B$ and $c_M$ parameters, using the $\aI = c_i \cdot a$ ({\bf left triangle plot}) and $\aI = c_i \cdot \Omega_{\rm DE}$ ({\bf right triangle plot}) parametrisations and different combinations of datasets and priors (see section \ref{sec-cosmo}).
All constraints shown assume $\aT = 0$, but note that constraints in the $c_M-c_B$ plane, as shown here, are only mildly affected by allowing $\aT \neq 0$ \cite{mcmc}.
Contours mark $68\%$ and $95\%$ confidence intervals, computed using just Planck data (P15), Planck data plus RSD, BAO and matter power spectrum measurements (P15 + LSS), and P15 + LSS with an additional prior ensuring the absence of GW-induced instabilities (P15 + LSS + GW). In fig. \ref{fig-1D} below we show that P15 + LSS + GW and P15 + GW lead to near identical constraints.
Dotted lines mark $c_i=0$ (the GR value), $c_B = 2$ (a singular line physical models cannot cross in the $\aI = c_i a$ parametrisation -- see \cite{Lagos:2017hdr,Lagos:2016wyv,mcmc} for details) and $c_M = -c_B$ (constraint derived from the GW prior).
\label{fig-comCon}}
\end{figure*}

\ni{\bf GW-induced decay and instabilities}:
Having applied the $\aT$ bound \eqref{ligo_bound}, \cite{Creminelli:2019kjy} recently argued that, in the presence of a sizeable cubic Horndeski operator, dark energy perturbations develop instabilities on gravitational wave backgrounds as sourced by massive black hole binaries. Requiring the absence of these induced gradient and ghost instabilities in terms of the $\aI$ then amounts to the following bound\footnote{Technically the bound of \cite{Creminelli:2019kjy} is that $\hat\aB \equiv -m_3^3/(2 \MPl^2 H) \lesssim 10^{-2}$, where $m_3^3$ is one of the free functions in the EFT action of \cite{Gleyzes:2013ooa} and we emphasise that their $\hat\aB$ differs from the standard $\aB$ of \cite{Bellini:2014fua}, as used here. Using the mapping of \cite{Bellini:2014fua}, in the absence of beyond-Horndeski interactions \cite{Zumalacarregui:2013pma,Gleyzes:2014dya}, we have $\tfrac{1}{2}M^2 (\aM + \aB) = \hat\aB \lesssim 10^{-2}$, 
so we are assuming $\tfrac{1}{2}M^2 \sim \Oo$ in \eqref{GWstab}. Since we will see that data constraints force $\aM$ to be small and positive (and hence $M^2$ will be at most slightly larger than unity at late times), this will be justified a posteriori.
}
\begin{align} \label{GWstab}
|\alpha_M + \alpha_B| \lesssim 10^{-2} \quad \Rightarrow \quad \aM \sim -\aB.
\end{align}
The implication is in the context of cosmologically observable dark energy-induced deviations from GR: Since next-generation experiments are expected to only be able to constrain the $\aI$ at the ${\cal O}(0.1)$ level \cite{Alonso:2016suf}, the condition |$\alpha_M + \alpha_B| \lesssim 10^{-2}$ implies that $\aM \sim -\aB$ for cosmologically significant $\aI$. From \eqref{alphadef2}, imposing this constraint within the subset of Horndeski theories \eqref{Horndeski_simple} amounts to suppressing the cosmological effects of the $G_3$ cubic Horndeski interaction at the $10^{-2}$ level \cite{Creminelli:2019kjy}. In other words, as far as cosmology is concerned the relevant theory for the evolution of large scale structure to leading order now becomes
\begin{align} \label{Horndeski_Vsimple}
{\cal L}_3 = \Lambda_2^4 G_2(\phi,X) + \MPl^2 G_4(\phi) R,
\end{align}
which implies that
\begin{align}
\aM &= -\aB = \frac{2\dot\phi}{H M^2} G_{4,\phi}, %&\aB &=-\aM,
\label{alphas}
\end{align}
as expected from \eqref{GWstab} and where $M^2 = 2 G_4$ and $\aT = 0$, as before. Once the background evolution is specified, $\aM$ is therefore the sole remaining free function relevant for linear perturbations in cosmology in this setup.

A brief note on extensions of Horndeski scalar-tensor theories. 
In so-called `beyond Horndeski' theories \cite{Zumalacarregui:2013pma,Gleyzes:2014dya} one additional $\aI$ enters at the level of linear perturbations, $\aH$. Similar considerations of GW-induced dark energy instabilities place highly restrictive bounds on this parameter, $|\aH| \lesssim 10^{-20}$ \cite{Creminelli:2019kjy}. Constraints of comparable strength are obtained from considering perturbative and resonant decay of gravitational waves into dark energy \cite{Creminelli:2018xsv,Creminelli:2019nok}. With bounds of this strength the additional free function becomes irrelevant for cosmology and we will ignore it (and the associated `beyond Horndeski' interactions) in what follows.  
%
%In principle also cubic Horndeski operator, but not at competitive level. 
%
Note that in even more general setups, so-called DHOST theories, \cite{Langlois:2015cwa,Crisostomi:2016czh}
computing associated constraints from GW speed and induced instability constraints introduces one additional class of interactions not constrained by current GW bounds \cite{Creminelli:2019kjy}. We will leave an exploration of the phenomenology associated with this additional freedom for future research. 
Finally note that, while the conformal $G_4(\phi)$ coupling can of course be absorbed by a conformal transformation, such a transformation does then change the coupling to matter (which cosmological constraints are also sensitive to). While both descriptions are physically equivalent, we here stay in Jordan frame, which is typically the more straightforward frame to use when comparing with large scale structure (LSS) data constraints.

\section{Cosmological parameter constraints} \label{sec-cosmo}

\begin{figure*}[t!]
\includegraphics[width=.49\linewidth]{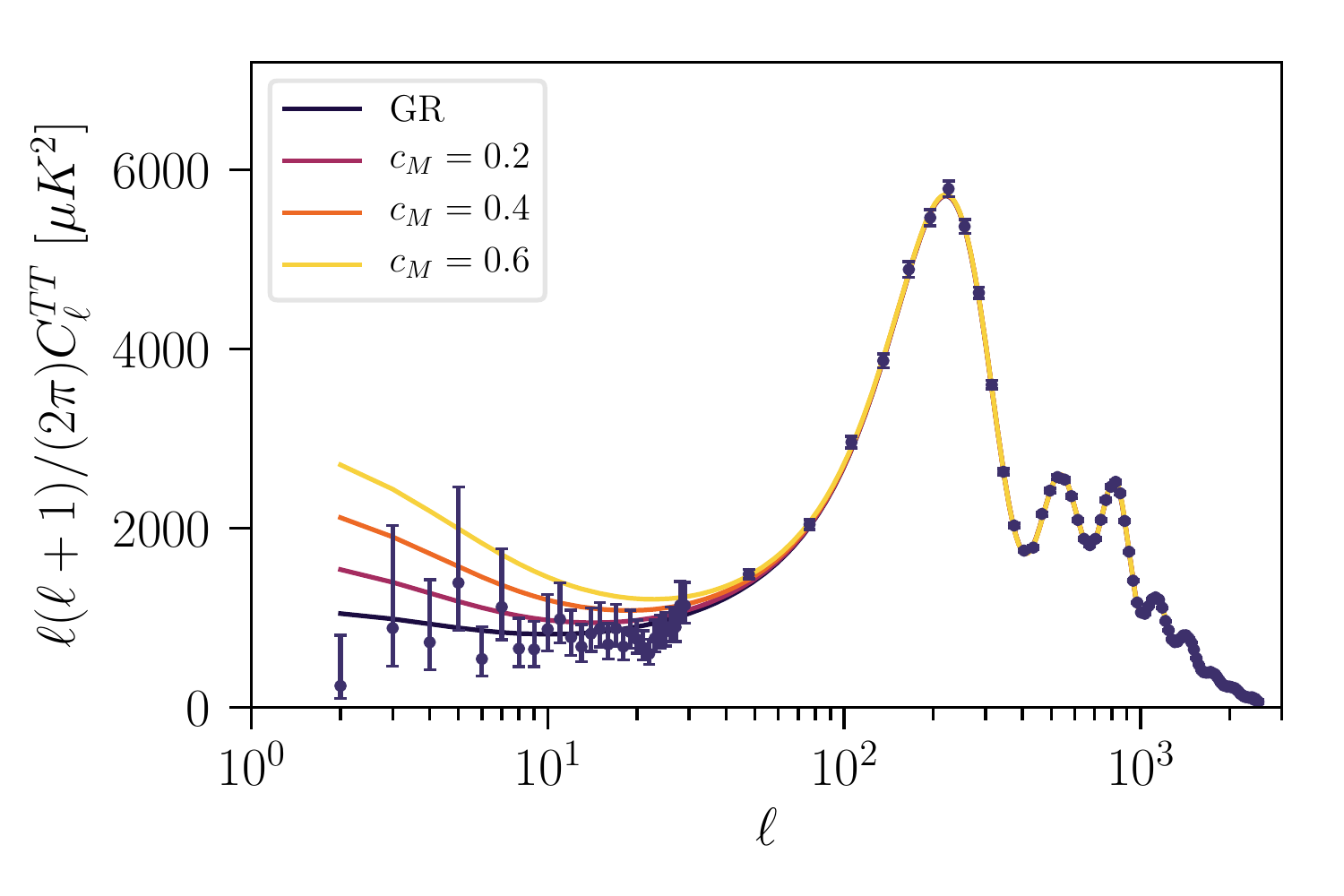}
\includegraphics[width=.49\linewidth]{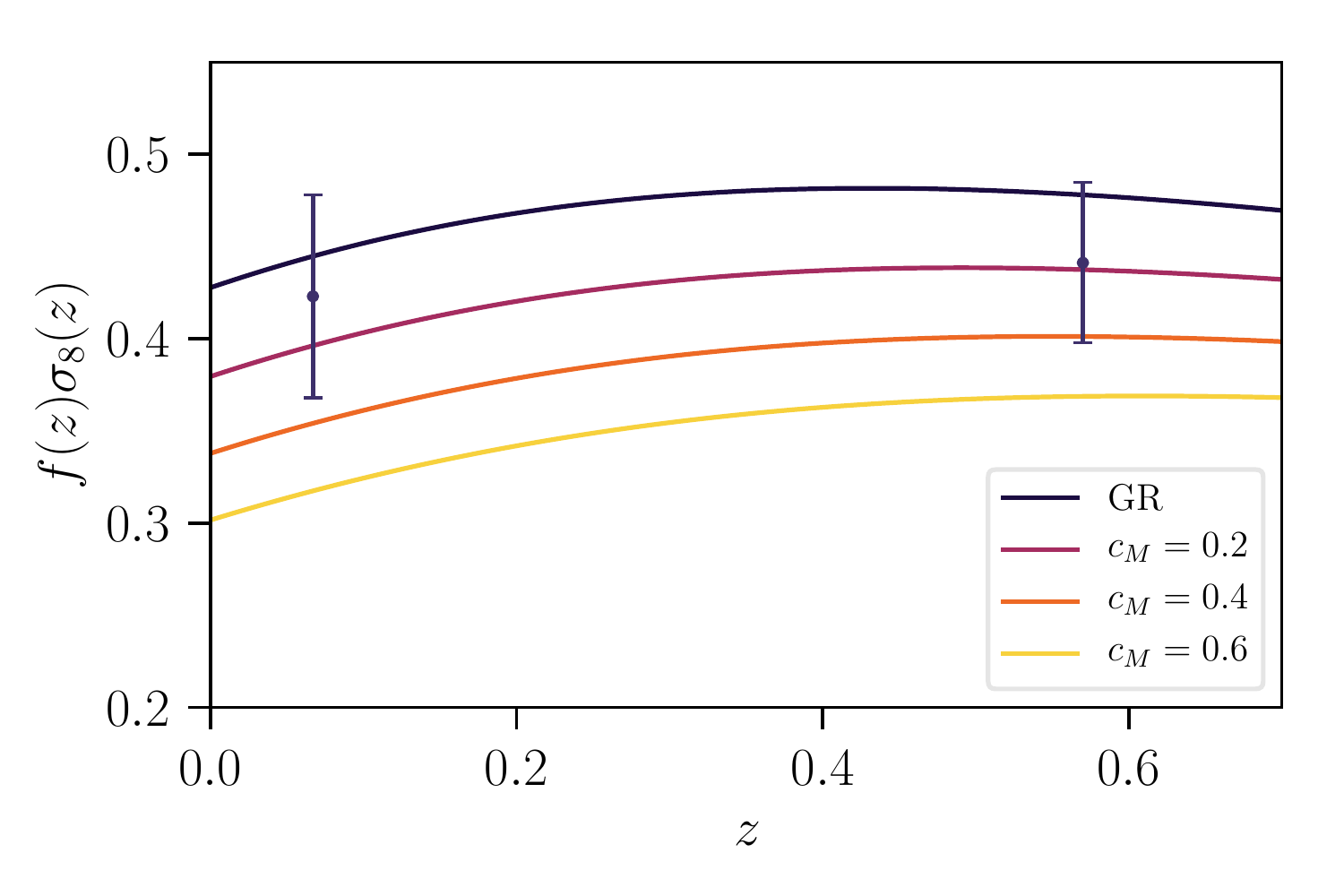}
\caption{
Illustration of the effect of varying $c_M = -c_B$ for the reduced Horndeski theory \eqref{Horndeski_Vsimple} consistent with the GW prior and when using the $\aI = c_i \cdot a$ parametrisation, on the CMB TT power spectrum ({\bf left plot}) and on $f\sigma_8$ ({\bf right plot}).
Data points are shown with $1\sigma$ uncertainties and all standard $\Lambda{}$CDM parameters are fixed to their Planck 2015 best-fit values here \cite{Planck-Collaboration:2016ae}. 
Increasing $c_M = -c_B$ for \eqref{Horndeski_Vsimple} very quickly adds too much power on large scales via the late ISW effect. In fact all $c_M \neq 0$ cosmologies shown here are ruled out at above $2\sigma$ via this effect alone and only $c_M < 0.06$ is consistent at this level -- c.f. fig. \ref{fig-comCon} and fig. \ref{fig-1D}. Planck data and the late ISW effect specifically therefore place the strongest bounds on \eqref{Horndeski_Vsimple}. 
$f\sigma_8$ constraints from RSDs, on the other hand, have strong constraining power for general Horndeski models (where they rule out large $c_M$ cosmologies consistent with CMB measurements, as long as $c_M$ and $c_B$ are both ${\cal O}(1)$ and positive), but only add very mild additional constraining power for the reduced Horndeski theory \eqref{Horndeski_Vsimple} (where $c_B$ is negative, whenever $c_M$ is positive).
\label{fig-TT}}
\end{figure*}

Having reduced the space of Horndeski theories to \eqref{Horndeski_Vsimple}, we will now compute cosmological data constraints on these theories and in particular on the sole remaining relevant free function at the level of linear perturbations, $\aM$.\footnote{For a comparison with cosmological parameter constraints on models without the $\aT$ constraint (and employing the same parametrisations as in this paper), see \cite{BelliniParam,mcmc}.} 
%\\
%
%
As discussed above, we consider such perturbations around an FRW background. More specifically we will assume this background to be $\Lambda${}CDM-like (motivated by the observed proximity to such a solution) and constrain perturbations around it. The background equations then read
\begin{align} \label{back}
H^2 &= \rho_{\rm tot},
&\dot H &= -\frac{3}{2}\left(\rho_{\rm tot}+p_{\rm tot}\right),
\end{align}
where $\rho_{\rm tot}$ and $p_{\rm tot}$ are the total energy density and pressure in the universe, and we have set $8\pi G = 1$ (and re-scaled all densities and pressures by a factor of 3, using CLASS conventions).
Computing cosmological constraints also requires choosing a parametrisation for the $\aI$ and this will turn out to be particularly important in the context of the reduced Horndeski theories \eqref{Horndeski_Vsimple} we are focusing on here. While numerous parametrisations exist -- for a discussion of their relative merits see Refs. \cite{Bellini:2014fua,BelliniParam,Linder:2015rcz,Linder:2016wqw,Denissenya:2018mqs,Lombriser:2018olq,Gleyzes:2017kpi,Alonso:2016suf,mcmc} -- here we will compute constraints using the two most commonly used ones \cite{Bellini:2014fua}: 
\begin{align} \label{oParam}
1) \quad \aI &= c_i a, &2) \quad \aI &= c_i \Omega_{\rm DE}.
\end{align}
These parameterise each $\aI$ in terms of just one constant parameter, $c_i$, and ensure that any deviation from GR (i.e. non-zero $\aI$) smoothly switches off towards higher redshift and only becomes relevant in the late universe.
\\

\ni{\bf Data sets and priors}:
We now perform a Markov chain Monte Carlo (MCMC) analysis, computing constraints on the modified gravity/dark energy parameters $c_M$ and $c_B$ for
%for the $\aT = 0$ Horndeski subset 
\eqref{Horndeski_simple} vs. just $c_M$ for 
%the reduced subset 
\eqref{Horndeski_Vsimple}, 
while marginalising over the standard $\Lambda{\rm CDM}$ parameters $\Omega_{\rm cdm}, \Omega_{\rm b}, \theta_s,A_s,n_s$ and $\tau_{\rm reio}$ -- for technical details regarding the MCMC implementation (as well as for additional details on the implementation and use of the data sets involved) see \cite{mcmc}. 
For related cosmological parameter constraints on deviations from GR using general parameterised approaches and a variety of (current and forecasted) experimental data, see
\cite{mcmc,BelliniParam,Hu:2013twa,Raveri:2014cka,Gleyzes:2015rua,Kreisch:2017uet,Zumalacarregui:2016pph,Alonso:2016suf,Arai:2017hxj,Frusciante:2018jzw,Reischke:2018ooh,Mancini:2018qtb,radstab,Perenon:2019dpc,Frusciante:2019xia,Arai:2019zul}
All the constraints computed here assume $\aT = 0$ as a prior to ensure compatibility with the speed of gravity constraints from GW170817 (as discussed in detail above), but we consider constraints with and without the further prior \eqref{GWstab} from requiring the absence of GW-induced gradient instabilities. For reference throughout the remainder of this section, we adopt the following shorthands in referring to the datasets we use (as well as the additional GW prior):
\begin{itemize}
\item {P15}: This prior includes Planck 2015 CMB temperature, CMB lensing and low-$\ell$ polarisation data \cite{Planck-Collaboration:2016af, Planck-Collaboration:2016aa, Planck-Collaboration:2016ae}.
\item {LSS}: Here we include baryon acoustic oscillation (BAO) measurements from SDSS/BOSS \cite{Anderson:2014, Ross:2015}, constraints from the SDSS DR4 LRG matter power spectrum shape \cite{Tegmark:2006} and redshift space distortion (RSD) constraints from BOSS and 6dF \cite{Beutler:2012, Samushia:2014}. Of these large scale structure measurements, RSDs have the strongest constraining power for general Horndeski theories \cite{mcmc}. 
\item {GW}:  This label denotes imposing the constraint \eqref{GWstab} as a prior, i.e. setting $\aM = -\aB$ (as far as large scale cosmology is concerned) to ensure the absence of GW-induced gradient instabilities. The residual difference of $\aM + \aB$ from zero at the $10^{-2}$ level is beyond the constraining power of current (see figure \ref{fig-comCon} and table \ref{tab_am_con}) as well as near-future CMB and LSS experiments \cite{Alonso:2016suf}. Unless explicitly noted, we also always impose $\aT = 0$, as discussed above.
\end{itemize} 
%\\
\vspace{.2cm}

\ni{\bf Data constraints}:
Results are shown and summarised in fig. \ref{fig-comCon} and table \ref{tab_am_con}. The improvement in constraining power in going from \eqref{Horndeski_simple} to \eqref{Horndeski_Vsimple}, i.e. when imposing the GW prior, is especially clear. While the prior of course eliminates one free function by design (we choose this to be $\aB$), constraints on the residual free function $\aM$ are also improved by an order of magnitude. 
The significant tightening of constraints for \eqref{Horndeski_Vsimple} is driven by two factors: 1) Gradient instabilities require $\aM \geq 0$. We will discuss the origin of this constraint in detail below. 2) Planck constraints rule out any $\aM$ above the $10^{-1}$ level.\footnote{This statement strictly applies for the $\aI = c_i a$ parametrisation. For the $\aI = c_i \Omega_{\rm DE}$ parametrisation, $\aM \lesssim 3 \cdot 10^{-1}$ -- see table \ref{tab_am_con}.} 
This happens, because the late-ISW effect is strongly enhanced as $\aM$ grows in size. Note that the GW prior $\aB \sim -\aM$ is crucial here: While modifications to the late ISW effect can be suppressed if $\aM$ and $\aB$ are enhanced simultaneously (this is the origin of the `degeneracy directions for P15-only constraints without the GW prior in fig. \ref{fig-comCon} -- see \cite{mcmc} for details), with the GW prior this is not possible and even small deviations away from GR lead to a strongly enhanced late-ISW effect. We show this explicitly in figure \ref{fig-TT}.

\begin{figure}[t]
\includegraphics[width=.49\linewidth]{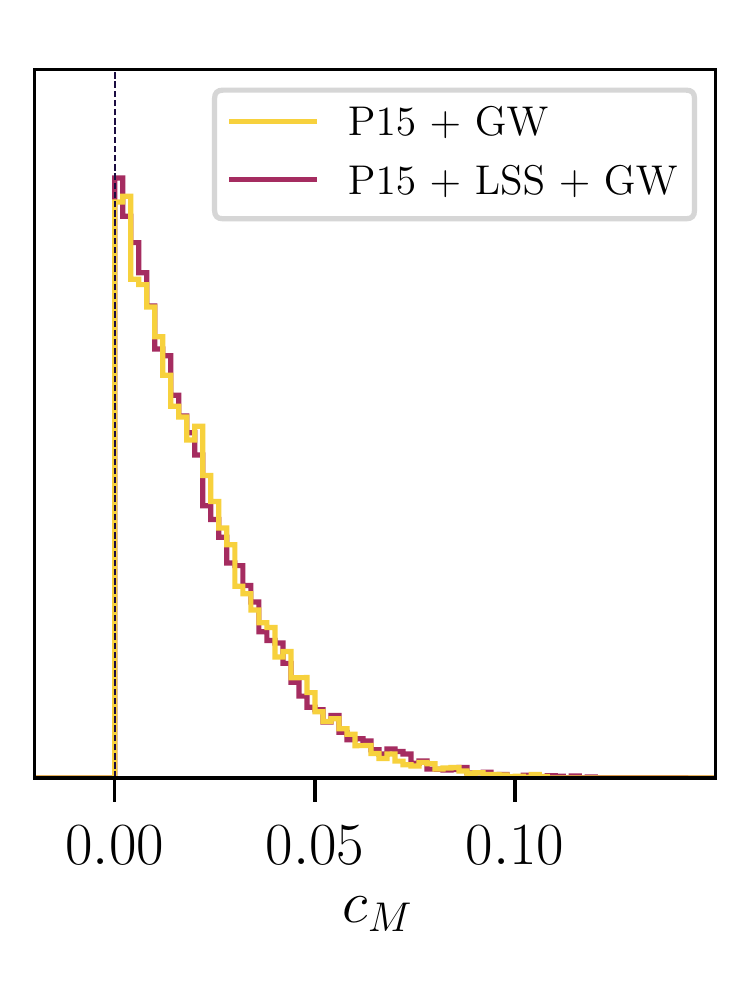}
\includegraphics[width=.49\linewidth]{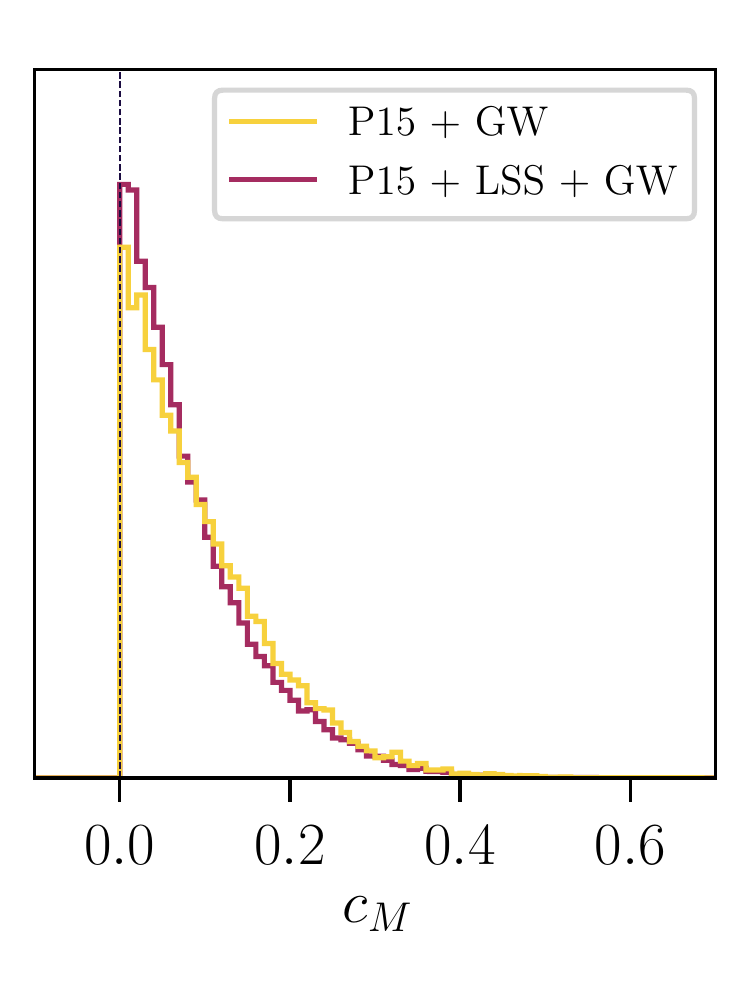}
%
%\end{center}
\caption{
The 1D posterior distribution of $c_M$, using the $\aI = c_i \cdot a$ ({\bf left plot}) and $\aI = c_i \cdot \Omega_{\rm DE}$ ({\bf right plot}) parametrisations and two different combinations of datasets and priors (see section \ref{sec-cosmo}). We compare constraints derived for the reduced Horndeski theory \eqref{Horndeski_Vsimple} (i.e. when imposing the GW prior) using Planck data with vs. without adding additional constraints from LSS measurements (as discussed in section \ref{sec-cosmo}). We find that, for this set of theories, constraints obtained with and without adding the above LSS data are near-identical.
\label{fig-1D}}
\end{figure}

In fig. \ref{fig-1D} we confirm that constraints on the $c_i$ are indeed driven by Planck data and largely independent of additional LSS measurements, whenever the GW prior is applied. 
Note that this agrees with the results shown in fig \ref{fig-TT}, which clearly shows that RSDs are less constraining than CMB $C_\ell^{TT}$ measurements for \eqref{Horndeski_Vsimple}, while they are a powerful additional constraint (ruling out large $c_M$) when the GW prior is {\it not} applied (see fig. \ref{fig-comCon}). 
Where there is overlap, our constraints agree well with the corresponding results of \cite{Ade:2015rim,radstab}, where (among other setups and for reasons unrelated to the GW priors discussed here) constraints for the reduced Horndeski subset \eqref{Horndeski_Vsimple} were computed using combinations of CMB and LSS data as well. \cite{Ade:2015rim} also explores adding different LSS data sets to their analysis, which only very mildly affects constraints. CMB measurements therefore dominate constraints, consistent with what we find here and as shown in fig. \ref{fig-1D}.

That constraints are driven by CMB data, whenever the GW prior is applied, in fact makes these constraints particularly robust: Including RSD and other LSS measurements in constraining deviations from requires careful theoretical and observational modeling that propagates and takes into account deviations from GR consistently at all levels, especially as related to scale-(in)dependent growth rates. For details we refer to \cite{Taruya:2013quf,Barreira:2016ovx,mcmc}, but emphasise that constraints for the reduced Horndeski set \eqref{Horndeski_Vsimple} presented here are only minimally affected by this issue (since data constraints are driven by CMB data).

Note that the strength of constraints for \eqref{Horndeski_Vsimple} computed here is in fact already comparable to what is expected (at least in the context of general Horndeski models) from (near-)future surveys via Fisher forecasts: In terms of the $\aI = c_i \Omega_{\rm DE}$ parametrisation also considered here, \cite{Alonso:2016suf} compute an expected $0.19$ uncertainty for $c_M$ at the $95\%$ confidence level\footnote{Notice that $c_M$ as defined here and in \cite{Alonso:2016suf} differ by a factor of $\Omega_{DE,0}$. The constraint quoted is mapping the result of \cite{Alonso:2016suf} to the convention used here.}, cf. fig. \ref{fig-1D} and table \ref{tab_am_con}. It would be interesting to re-visit such forecasts while incorporating the priors considered here, since this may result in smaller projected uncertainties for $c_M$. Finally, note that, when comparing constraints on $c_M$ for the $\aI = c_i a$ vs. 
$\aI = c_i \Omega_{\rm DE}$ parametrisation (cf. figure \ref{fig-1D}), constraints are tighter for the former, since it has a stronger effect at earlier times (and hence larger $\ell$ in figure \ref{fig-TT}), when CMB bounds are particularly constraining.
\\

\section{Stability and parametrisation dependence} \label{sec-stab}

In the context of \eqref{Horndeski_Vsimple}, in the previous section we discussed the crucial nature of CMB constraints in ruling out $c_M \gtrsim {\cal O}(0.1)$. Requiring the absence of gradient instabilities provides the complementary bound ruling out $c_M < 0$, which then (together with CMB bounds) results in the very tightly constrained $c_M$. Here we discuss the precise nature of these instabilities and to what extent their presence/absence (and resulting effect on constraints) is robust to the use of different parametrisations for the $\aI$. %such as \eqref{oParam}.
\\

\ni{\bf Stability}:
Gradient instabilities are sourced when the effective speed of sound $c_s$ becomes imaginary and affect the large $k$ (i.e. sub-horizon and high energy) regime. Such instabilities generically lead to an uncontrolled growth of perturbations that invalidates the associated cosmological solution.
For a general Horndeski theory \eqref{Horndeski_action} the speed of sound $c_s$ satisfies
\begin{align} \label{sound_speed}
{\cal D} c_s^2 = (2-\aB)\left(\hat\alpha - \frac{\dot H}{H^2}\right) - \frac{3(\rho_{\rm tot} + p_{\rm tot})}{H^2M^2} + \frac{\dot{\alpha}_B}{H},
\end{align}
where  we have defined $\hat \alpha \equiv \tfrac{1}{2}\aB(1+\aT) + \aM - \aT$ and ${\cal D} \equiv \aK + \tfrac{3}{2}\aB^2$. ${\cal D} $ has to be positive by virtue of the no-ghost requirement for physical scalar perturbations.

When specialising to theories with $\aT =0$, i.e. to \eqref{Horndeski_simple} in the present context, requiring the absence of gradient instabilities therefore amount to imposing
\begin{align} \label{gradient_condition2}
(2-\aB)\left(\tfrac{1}{2}\aB + \aM - \frac{\dot H}{H^2}\right) - \frac{3(\rho_{\rm tot} + p_{\rm tot})}{H^2M^2} + \frac{\dot{\alpha}_B}{H} \geq 0. 
\end{align}
In terms of constraints, the lower left boundary of the contours in fig. \ref{fig-comCon} corresponds to the onset of such instabilities and gradient instabilities, e.g. ruling out most cosmologies with $\aM < -\aB$. 
Applying the GW prior, so setting $\aM = -\aB$, then further reduces the above stability condition to  
\begin{align} \label{gradient_condition3}
(2+\aM)\left(\tfrac{1}{2}\aM - \frac{\dot H}{H^2}\right) - \frac{3(\rho_{\rm tot} + p_{\rm tot})}{H^2M^2} - \frac{\dot{\alpha}_M}{H} \geq 0.
\end{align}
It will also be instructive to evaluate this on the background ($\Lambda$CDM) equations of motion \eqref{back}, phrasing everything in terms of $M^2, H$ and their time derivatives and resulting in
\begin{align} \label{gradient_condition31}
\aM(1+\tfrac{1}{2}\aM) - \frac{\tfrac{d}{dt}(\aM H)}{H^2} - 2\frac{\dot H}{H^2} \frac{(M^2 - 1)}{M^2} \geq 0.
\end{align}

We are now in a position to obtain a better understanding of what controls the onset of such instabilities, by evaluating the above condition for a given parametrisation. We first focus on a generalisation of the $\aM = c_M a$ parametrisation, namely a power law $\aM = c_M a^p$. In this case we obtain $\dot \aM/H = p \aM$, so that the gradient stability condition simplifies to
\begin{align} \label{gradient_condition4}
\aM \left(1 - p - \frac{\dot H}{H^2}\right) + \tfrac{1}{2}\aM^2 - 2\frac{\dot H}{H^2} \frac{(M^2 - 1)}{M^2} \geq 0.
\end{align}
Given that the above data constraints require $\aM \ll 1$, we can generically drop the $\aM^2$ term to leading order, but will keep it in place for the time being (it will play a role during dark energy domination). Evaluating \eqref{gradient_condition4} during a de Sitter phase, matter and radiation domination, we find the following stability conditions
\begin{align} 
{\rm dS} &: \quad\quad \aM \left(1 - p\right) + \tfrac{1}{2}\aM^2 \geq 0, \nn \\
{\rm mat} &: \quad\quad \aM \left(\tfrac{5}{2} - p\right) + \tfrac{1}{2}\aM^2 +3 \frac{(M^2 - 1)}{M^2} \geq 0, \nn \\
{\rm rad} &: \quad\quad \aM \left(3 - p\right) + \tfrac{1}{2}\aM^2 + 4 \frac{(M^2 - 1)}{M^2} \geq 0.
\label{dS_mat_rad_v1}
\end{align}
where we have used that, from \eqref{back}, $\dot H/H^2$ takes the values of $0, -3/2$ and $-2$ during a de Sitter phase, matter and radiation domination, respectively.
For $p = 1$, i.e. our main $\aM = c_M a$ parametrisation, the evolution is stable for all positive $\aM$. All individual terms in \eqref{dS_mat_rad_v1} are positive (or zero) then, since $M^2  > 1$ for positive $\aM$. Note that the de Sitter condition simply becomes $\aM^2 \geq 0$, so that condition is satisfied independently of the sign of $\aM$. The matter and radiation domination conditions, on the other hand, are violated for negative $\aM$, since the two dominant $M^2$ and $\aM$ terms contribute with negative sign then (the conclusion remains true for large negative $\aM$ as well). These instabilities therefore rule out negative $\aM$, as expected. For $p > 1$ and small, positive $\aM$, instabilities generically develop in the de Sitter phase, so such parametrisations will likely be severely constrained/ruled out by a combination of the de Sitter gradient stability condition $\aM \gtrsim 1$ and ISW constraints from the CMB (which will prefer a small $\aM$). We leave a more detailed exploration of such scenarios for future work. 
\\

\begin{table*}[t]
\begin{tabular}{cc}
    \begin{minipage}{.5\linewidth}
\renewcommand{\arraystretch}{1.8}
\setlength{\tabcolsep}{0.3cm}
\begin{tabular}{|l||cc|} \hline
{$\aM = c_M \cdot a$} & ${\rm P15}$ & ${\rm P15} + {\rm LSS}$\\ \hline\hline
GW17 & $0^* \leq c_m \leq 1.48^*$ & $0^* \leq c_m < 0.71 $\\ \hline
GW17 + GW19 & $0^* \leq c_m < 0.06$ & $0^* \leq c_m < 0.06$\\ \hline
\end{tabular}
    \end{minipage} &

    \begin{minipage}{.5\linewidth}
\renewcommand{\arraystretch}{1.8}
\setlength{\tabcolsep}{0.3cm}
\begin{tabular}{|l||cc|} \hline
{$\aM = c_M \cdot \Omega_{\rm DE}$} & ${\rm P15}$ & ${\rm P15} + {\rm LSS}$\\ \hline\hline
GW17 & $1.48\substack{+3.21 \\ -1.91}$ & $0.20\substack{+1.18 \\ -0.82}$\\ \hline
GW17 + GW19 & $0^* \leq c_m < 0.27$ & $0^* \leq c_m < 0.25$\\ \hline
\end{tabular}
    \end{minipage} 
\end{tabular}
\caption{Posteriors on the dark energy $c_M$ parameter, for the $\aM = c_M \cdot a$ ({\bf left table}) and $\aM = c_M \cdot \Omega_{\rm DE}$ ({\bf right table}) parametrisations and different combinations of datasets and priors (see section \ref{sec-cosmo}). Note that we here distinguish between two GW priors: GW17 denotes setting $\aT = 0$ (motivated by constraints from GW170817) and GW19 denotes setting $\aM = -\aB$ (motivated by requiring the absence of GW-induced instabilities).
Uncertainties shown denote the $95 \%$ confidence level. When the $c_M$ distribution is strongly non-Gaussian (as is e.g. always the case when the GW19 prior is applied), we do not give a mean value and denote limit values due to prior boundaries (when there is an excellent fit to the data on that boundary) with an asterisk. 
}
\label{tab_am_con}
\end{table*}

\ni{\bf Parametrisation dependence}: Above we discussed the gradient stability condition and its effect on constraints in the context of the $\aM = c_m a$ parametrisation (and power-law generalisations thereof). However, the conclusions drawn from this analysis are somewhat parametrisation-dependent and we will see that this is especially important in the context of the subset of theories \eqref{Horndeski_Vsimple} consistent with the GW prior. To make this more explicit, consider the $\aM = c_m \Omega_{\rm DE}$ parametrisation. 
We note that $\dot \aM/H = -2 (\dot H/H^2) \aM$ in this parametrisation and evaluate \eqref{gradient_condition31} accordingly, finding
%For oParam we on the other hand  so that the gradient stability condition becomes
\begin{align} \label{gradient_condition_5}
\aM \left(1 + \frac{\dot H}{H^2}\right) + \tfrac{1}{2}\aM^2 - 2\frac{\dot H}{H^2} \frac{(M^2 - 1)}{M^2} \geq 0.
\end{align}
Here again the $\aM^2$ term is negligible at leading order (for observationally relevant small $\aM$), but it is clear that the first term already yields a negative contribution during matter domination. Making this more explicit, we can again evaluate \eqref{gradient_condition_5} for different regimes
\begin{align} 
{\rm dS} &: \quad\quad \aM + \tfrac{1}{2}\aM^2 \geq 0, \nn \\
{\rm mat} &: \quad\quad -\tfrac{1}{2} \aM + \tfrac{1}{2}\aM^2 +3 \frac{(M^2 - 1)}{M^2} \geq 0, \nn \\
{\rm rad} &: \quad\quad -\aM + \tfrac{1}{2}\aM^2 + 4 \frac{(M^2 - 1)}{M^2} \geq 0.
\end{align}
The de Sitter constraint now by itself eliminates negative $\aM$ (for small $\aM$). However, for positive $\aM$ and during matter and radiation domination there is a tension between the negative linear $\aM$ term and the positive contribution of the $M^2$-dependent term. 
For sizeable, yet observationally allowed $c_M$ (i.e. $c_M$ of order $10^{-1}$), there is no instability during matter domination and the instability appears at redshifts above (i.e. higher than) matter-radiation equality.  Only for highly suppressed $\aM$, more specifically $c_M \lesssim 10^{-3}$, this instability already appears at redshifts before (i.e. lower than) matter-radiation equality. However, in both cases the instability is present in the radiation-dominated era, so strictly applying a gradient stability prior would exclude all non-GR cosmologies with the GW prior in this parametrisation. 

In appendix \ref{appendix_grad} we discuss whether such instabilities are catastrophic for the model/parametrisation considered or not. The conclusion depends on a number of technical details, but it is important to not lose sight of the overall picture. 
As discussed in the appendix, modifying the $\aM = c_M \Omega_{\rm DE}$ parametrisation with the addition of a sufficiently enhanced early-time $\aK$ suppresses gradient instabililities during the radiation-dominated era to the extent that these no longer affect relevant phenomenology. 
Constraints shown for the $\aM = c_M \Omega_{\rm DE}$ parametrisation in section \ref{sec-cosmo} follow this approach (following previous literature \cite{mcmc,radstab} and setting $c_k = 0.1$ and $d_k = 10^{-2}$, in the notation of appendix \ref{appendix_grad}).
Whether any such hybrid parametrisation 
is born out by realistic models is beyond the scope of this paper, but a more conservative view is that parametrisations should not be trusted in regimes where they lead to (gradient or indeed ghost) instabilities. From this point of view the `failure' of the $\aM = c_M \Omega_{\rm DE}$ parametrisation in the radiation-dominated era is not surprising -- this as well as other parametrisations are primarily designed to capture the leading-order behaviour of dark energy perturbations in the regime where they are most relevant, i.e. during the very late universe. In this regime they are a good tool to obtain reasonably accurate and model-independent estimates of constraints on dark energy, but simple one parameter models are of course at best a leading order approximation and cannot reasonably be expected to accurately track the physical evolution of dark energy perturbations across different regimes. So while e.g. enhancing $\aK$ at early times to suppress instabilities is indeed a `hack', it may be seen as a proxy to recover an approximate early time dark energy evolution of more realistic models. Ultimately such parametrisations will of course be justified or rejected a posteriori, based on whether they do indeed recover the evolution of physically motivated models.
At the current and still early stage in the exploration of parameter space it is therefore perhaps too early to exclude a hybrid parametrisation that removes early universe instabilities by fiat (as discussed above) on essentially aesthetic grounds. 
We leave a more detailed analysis of such parametrisations for future research, but hope that the aspects discussed here will help guide the construction of increasingly theoretically informed and realistic parametrisations. 
\\

\section{Strong coupling scale and screening} \label{sec-screen}

In addition to cosmological constraints for \eqref{Horndeski_Vsimple}, one immediate other concern is how consistency with fifth force constraints is ensured, especially with very tight bounds on solar system scales (e.g. from lunar laser ranging \cite{Williams:2004qba,Merkowitz:2010kka}). This concern at least partially stems from the disappearance of all higher derivative interactions suppressed by $\Lambda_3$ in \eqref{Horndeski_Vsimple}, seemingly removing such interactions as potential sources of Vainshtein screening \cite{Vainshtein:1972sx} (for general reviews see e.g. \cite{Khoury:2010xi,Babichev:2013usa}). While Vainshtein screening may still be present to an extent and associated with higher-derivative interactions inside $G_2$ (e.g. of the k-essence $X^n$ type \cite{ArmendarizPicon:1999rj,Garriga:1999vw,ArmendarizPicon:2000dh,ArmendarizPicon:2000ah} -- see \cite{Babichev:2009ee,Brax:2012jr,Burrage:2014uwa,deRham:2014wfa} for screening-related discussions in this setup) and Chameleon screening \cite{Khoury:2003aq,Khoury:2003rn} can be associated with the conformal $f(\phi)$ coupling\footnote{Transforming \eqref{Horndeski_Vsimple} to the Einstein frame involves sending $g_{\mu\nu} \to G_4^{-1}(\phi) g_{\mu\nu}$, so the conformal $G_4(\phi)$ coupling is mapped into a conformal factor in the effective metric that matter couples to, as is characteristic for Chameleon screening. Note that the conformal nature of this coupling is essential -- disformal contributions cannot contribute to this type of screening at leading order \cite{Noller:2012sv}.} 
(see \cite{Burrage:2017qrf} for a review of Chameleon-related constraints), 
here instead we wish to explore another option: Following \cite{Noller:2019chl} we want to understand the above GW priors (on the speed of GWs and on the absence of GW-induced dark energy instabilities) as constraints on the strong coupling scale $\Lambda$ of general Horndeski theories and understand what this implies for that scale and for related screening mechanisms. 
\\

\ni {\bf Strong coupling scale and GW constraints}:
Phrased in terms of the $\aI$ parameters, the two GW priors we have discussed throughout this paper are $|\aT| \lesssim 10^{-15}$ and $|\aM + \aB| \lesssim 10^{-2}$. Instead of simply setting the associated (pieces of the) $G_3, G_4$ and $G_5$ interactions to zero, these constraints can also be interpreted as bounds on the scale $\Lambda$ suppressing the higher derivative interactions contributing to $\aT$ and/or $\aM + \aB$. In \eqref{Horndeski_lagrangians} this was set to $\Lambda = \Lambda_3$, which ensured that interactions suppressed by $\Lambda = \Lambda_3$ can still yield $\Oo$ contributions to the background evolution. When raising $\Lambda$, contributions from the associated operators will generically become suppressed at the cosmological level, but may still contribute at smaller scales. For example, in \cite{Noller:2019chl} it was shown that constraints on $\aT$ can be satisfied in this way, while still yielding observable $\Oo$ modifications around black hole space times.

When freeing up $\Lambda$ in this way, one should re-visit \eqref{Horndeski_lagrangians}, where we had used the identity $\MPl^2 = \Lambda_2^8/\Lambda_3^6$ to re-write all scales in terms of $\Lambda_2$ and $\Lambda_3$. Focusing on ${\cal L}_4$ for the time being, it is useful to think of $G_4(\phi,X)$ as being expanded in powers of $\phi$ and $X$. All $X$-dependent pieces contribute to the $G_{4,X}$ term, so when the scale of this is raised, the scale of the corresponding piece of the $G_4 R$ interaction needs to be raised accordingly in order for the Horndeski tuning to be maintained. The $X$-independent, i.e. solely $\phi$-dependent or constant pieces of $G_4$ are unaffected by $\Lambda$, so are oblivious to any change in this scale. It is therefore instructive to re-write ${\cal L}_4$ in \eqref{Horndeski_lagrangians} as
\begin{align}
{\cal L}_{4}  & = \MPl^2 (\tfrac{1}{2} + \tilde G_4) R + \frac{\Lambda_2^8}{\Lambda^6} \hat G_{4} R + \frac{\Lambda_2^4}{\Lambda^6} ~ \hat G_{4,X} \left( [\Phi]^2-[\Phi^2] \right),
\end{align}
where we have separated out a canonical kinetic term for the graviton $\tfrac{1}{2}\MPl^2 R$, and write $G_4 = \tilde G_4 + \hat G_4$. Here $\hat G_4$ includes all $X$-dependent contributions, whereas $\tilde G_4$ collects all purely $\phi$-dependent contributions. Any scale suppressing interactions in $\tilde G$ is therefore decoupled from $\Lambda$, so the effects of a conformal $\tilde G$ (or $G_4(\phi)$) coupling are not suppressed when raising $\Lambda$.\footnote{Note that $\tilde G$ vanishes in shift-symmetric setups, which are particularly well-motivated from the point of view of radiative corrections \cite{Pirtskhalava:2015nla}, but reduce \eqref{Horndeski_Vsimple} to a k-essence theory. So we will not insist on shift symmetry here.}
Analogous considerations apply to the $G_5$ interactions, with the simplifying exceptions that any constant contribution vanishes (up to boundary terms) and that any purely $G_5(\phi)$ interaction can be integrated-by-parts and absorbed into lower order $G_i$ interactions, so only the $X$-dependent pieces of $G_5$ remain, scaling as $1/\Lambda^9$. 

Going back to the GW constraints on $\aT$ and $\aM + \aB$, we can now simply read off how these scale when $\Lambda$ is raised, finding
\begin{align}
\aM + \aB &\sim \left(\frac{\Lambda_3}{\Lambda}\right)^3, &\aT &\sim \left( \frac{\Lambda_3}{\Lambda} \right)^6,
\end{align}
where we keep the lowest powers of $\Lambda_3/\Lambda$.\footnote{Once $|\aT| \lesssim 10^{-15}$ has been achieved by raising $\Lambda$ or by some other tuning, the scaling of $\aM + \aB$ can also be lifted from the resulting expression
\begin{align}
\aM + \aB = - 2\frac{\dot{\phi}X}{H M^2} G_{3,X}.
\end{align}
}
The $\aT$ constraint can therefore be satisfied by setting $\Lambda \gtrsim 10^3 \Lambda_3$ \cite{Noller:2019chl}, where in the EFT-spirit we implicitly assume that all other dimensionless coefficients in the theory are $\Oo$. Raising $\Lambda$ at this level also automatically yields $|\aM + \aB| \sim 10^{-9}$, very comfortably satisfying the $10^{-2}$ level constraint from requiring the absence of GW-induced dark energy instabilities. In fact, if the $\aT$ constraint has already been satisfied by some other tuning -- which eliminates all contributions from $G_{4,X}$ and $G_5$, so our parent theory now is \eqref{Horndeski_simple} -- then raising $\Lambda$ by just one order of magnitude, i.e.  $\Lambda \gtrsim 10 \; \Lambda_3$, already leads to $|\aM + \aB| \sim 10^{-3}$ in agreement with all GW priors. 
\\

\ni {\bf Vainshtein screening}:
The above considerations immediately suggest that $\Lambda$ can still play a significant role in satisfying fifth force constraints.
While the non-linear interactions suppressed by $\Lambda$ no longer lead to $\Oo$ effects on cosmological scales, they can still lead to Vainshtein screening \cite{Vainshtein:1972sx} on smaller scales. 
To this end consider the Vainshtein radius $r_V$ in a spherically symmetric and static setup around an object of mass $M$. Inside this radius non-linearities in the Horndeski scalar dominate and (upon canonical normalisation) the scalar effectively decouples from matter, suppressing the mediation of a fifth force. The Vainshtein radius $r_V$ associated with Galilean symmetric \cite{Nicolis:2008in} part of the interactions suppressed by $\Lambda$ is given by \cite{Koyama:2013paa}
\begin{align} \label{rV}
r_V = \frac{1}{\Lambda}\left(\frac{M}{8\pi \MPl}\right)^{\frac{1}{3}},
\end{align}
where we assume a gravitational strength coupling of matter and the scalar in the original theory.\footnote{Note that, if this coupling is not of gravitational strength, but modified by a dimensionless coupling constant $\delta$, this will enter in \eqref{rV} as an extra $\delta^{1/3}$ factor \cite{Noller:2019chl}. If the coupling is indeed only generated via de-mixing the $\aB$-induced scalar-tensor mixing of cosmological perturbations, then we have $\delta \sim \aB$. For $\aM = -\aB \gtrsim 10^{-2}$, i.e. the cases of phenomenological interest discussed throughout where these $\aI$ are primarily generated by $G_4(\phi)$, this factor then is $\delta^{1/3} \gtrsim 0.2$. In other words, how precisely the coupling is generated does not qualitatively affect the estimates of $r_V$ here.}
When $\Lambda = \Lambda_3$, as was the case for \eqref{Horndeski_lagrangians}, the Vainshtein radius for the Sun is $0.1 \; {\rm kpc} \sim 10^{15} \; \km$ (for comparison, the radius of the Milky Way is $\sim 10^{17} \; \km$), while the size of the solar system is $\sim 10^{9} \; \km$.
\footnote{This is the approximate distance for Jupiter/Saturn, while Neptune is at a distance of $\sim 4.5 \cdot 10^9 \; \km$ from the Sun.} 
The solar system is therefore comfortably inside the Vainshtein radius and the scalar is screened on solar system scales. Freeing $\Lambda$, we therefore have
\begin{align}
r_V \sim \left(\frac{\Lambda_3}{\Lambda}\right) \cdot \left(\frac{M}{M_{\odot}}\right)^{\frac{1}{3}} \cdot 10^{15} \; \km,
\end{align}
where $M_{\odot}$ is the mass of the sun.
Raising $\Lambda$ by one order of magnitude (assuming $\aT = 0$ is ensured by some other tuning) leaves a Vainshtein radius (for the Sun) at $r_V \sim 10^{14} \; \km$, i.e. much larger than the size of the solar system. 
Similarly, raising $\Lambda$ by three orders of magnitude in order to both satisfy the GW speed \eqref{ligo_bound} and stability \eqref{GWstab} bounds still leaves a Vainshtein radius (for the Sun) at $r_V \sim 10^{12} \; \km$, comfortably leaving the solar system Vainshtein screened.\footnote{ 
If we also take into account the `beyond positivity' bound of \cite{Bellazzini:2019xts} (also see \cite{deRham:2017imi,Bellazzini:2017fep,deRham:2017xox}), derived by assuming properties of the theory's UV completion, then the lower limit on $\Lambda$ may be raised yet again. In the context of weakly broken Galileons/shift symmetric Horndeski models \cite{Pirtskhalava:2015nla}, this implies that $\Lambda_{\rm UV}$, the scale where the regime of validity of the (cosmological) EFT ends, satisfies $\Lambda_{\rm UV} \lesssim 10^{-7} (\Lambda/\Lambda_3)^{3/2} \; \km^{-1}$ \cite{Bellazzini:2019xts}. If $\Lambda \sim 10 \Lambda_3$, the EFT then already breaks down on scales of $10^{5} \; \km$, so similar to the earth-moon orbit. In this context at least some solar system tests would therefore not even test the cosmological EFT anymore, but depend on the precise properties of the UV completion. If instead one insists on solar system tests as well as strong field tests around black holes to be within the regime of validity of the cosmological EFT, one should require $\Lambda_{\rm UV} \lesssim 1 \; \km^{-1}$, translating into $\Lambda \gtrsim 10^5 \Lambda_3$ \cite{Noller:2019chl}. This would of course automatically satisfy both GW priors, but also bring down the previously discussed Vainshtein radius to $\sim 10^{10} \; \km$, i.e. very close to solar system scales. Depending on restrictions placed on $\Lambda_{\rm UV}$, a Vainshtein screening mechanism solely based on $\Lambda$-suppressed interactions with $\Lambda \gtrsim 10^5 \Lambda_3$ might therefore leave observable unscreened fifth force effects in the solar system in tension with current (observational) constraints. We will leave this topic for future work and remain agnostic about properties of the UV completion here, only enforcing the (low-energy) bounds we have discussed so far.}
While the effect of a reduced $r_V$ on larger (galactic) scales clearly needs to be explored further, $\Lambda$ can therefore be raised to address both GW priors, while also providing a functional screening mechanism to deal with the especially tight and well-understood fifth force constraints on solar system scales. Of course none of these $\Lambda$-related considerations constrain \eqref{Horndeski_Vsimple}, so the non-trivial cosmological phenomenology discussed in the previous sections is unaffected by this.
\\

\section{The quasi-static limit} \label{sec-qsa}
When focusing on intermediate scales within the sound horizon  $c_s^2k^2/a^2\gg H^2$ (i.e. in practice for the majority of scales we have observational access to, at least as long as $c_s^2 \sim 1$), the quasi-static approximation (QSA) closely recovers the full evolution of gravitational perturbations \cite{Sawicki:2015zya}.\footnote{In the linear ST regime considered here, the accuracy of the QSA can also be linked to the proximity of the background to $\Lambda{}$CDM \cite{delaCruzDombriz:2008cp,Noller:2013wca}.}
We therefore here wish to explore what the GW prior and the above data constraints imply in the QSA. 
\\

\ni {\bf The quasi-static approximation}: The QSA amounts to assuming $|\dot{X}| \sim {H} |X|\ll |\partial_i X|$ for any gravitational perturbation field $X$, and hence to a sub-horizon regime where time derivatives of gravitational perturbations can be neglected in comparison to their spatial derivatives. For scalar perturbations (in the standard scalar-vector-tensor decomposition \cite{Kodama:1985bj}) in Newtonian gauge we then have 
\begin{equation}\label{Def4Pert}
ds^2=-\left(1+2\Phi\right)dt^2+ a^2\left(1-2\Psi\right)dx_i dx^i.
\end{equation}
The effective Poisson equation as well as the relation between the metric Bardeen potentials (as related to anisotropic stress and lensing observables) can then be parametrised as follows
\begin{align}
\frac{k^2}{a^2}\Phi &=-4\pi G\mu(a) \rho\Delta, \nn \\
\gamma(a) &=\frac{\Psi}{\Phi}, \nn \\
%
%\frac{k^2}{a^2}\left(\Phi+\Psi\right) &=-8\pi G \,\Sigma(a)\left[\rho \Delta+\frac{3}{2}(\rho+P)%\sigma_m\right],
\frac{k^2}{a^2}\left(\Phi+\Psi\right) &=-8\pi G \,\Sigma(a)\rho \Delta,
\label{mrs}
\end{align}
where $\Delta$ is the comoving gauge invariant density perturbation,
$G$ is Newton's constant satisfying $G=1/(8\pi M_P^2)$, 
and we have ignored any anisotropic stress source in the matter sector.
Here $\mu$ quantifies modifications to the effective strength of gravity (i.e. an effective different Newton's constant), $\gamma$ measures the presence of an effective gravitational anisotropic stress (often called gravitational ``slip'') and $\Sigma$ parametrises modifications to the effective lensing potential probed by gravitational lensing. GR corresponds to the case when $\mu,\gamma,\Sigma$ are all unity, so deviations away from unity for any of these parameters are probing the presence of new gravitational \dof.
From \eqref{mrs} one can see that the three parameters $\mu,\gamma,\Sigma$ are not independent, but instead are related by
\begin{equation}
\Sigma=\frac{1}{2}(1+\gamma)\mu.
\label{sigmugam}
\end{equation}

In evaluating the QSA parameters \eqref{mrs} for the highly restricted Horndeski theory \eqref{Horndeski_Vsimple}, we will find it useful to follow \cite{Alonso:2016suf,Lagos:2017hdr} and introduce a number of $\hat\beta_{i}$ functions as a shorthand
\begin{align}
\beta_1 &\equiv - \frac{3(\rho_{\rm tot} + p_{\rm tot})}{H^2M^2} - 2\frac{\dot{H}}{H^2} + \frac{\tfrac{d}{dt}{(\aB H)}}{H^2}, \nn \\
\beta_2 &\equiv \aB (1+\aT) + 2(\aM - \aT), \nn \\
\beta_3 &\equiv (1+\aT)\beta_1 + (1+\aM)\beta_2, \nn \\
\beta_4 &\equiv \aB (\aT - \aM) - \tfrac{1}{2}\aB^2(1 + \aT).
\label{betas}
\end{align}  %
The first three functions are as defined in \cite{Alonso:2016suf,Lagos:2017hdr} and we add $\beta_4$ here. With these definitions we can write the expression for the speed of sound $c_s^2$ \eqref{sound_speed} in the following compact form
\begin{align}
{\cal D} c_s^2 = \beta_1 + \beta_2 + \beta_4. 
\end{align}
We can similarly express the quasi-static parameters $\gamma$ and $\mu$ in terms of the $\hat\beta_i$ functions, finding
\begin{align}
M^2\mu &= \frac{2\beta_3}{2\beta_1+\beta_2(2-\aB)}, \nn \\
\gamma \;\; &= %\quad\quad 
\frac{\beta_1+\beta_2}{\beta_3} \quad\quad \; %= \frac{\hat\beta_1+\aM}{\beta_1 +\aM(1+\aM)},
\label{mugamma}
\end{align}
where $M^2$ is the effective Planck mass, as before, and we will find it useful to define the shorthands $\hat \mu \equiv M^2 \mu$ and $\hat \Sigma \equiv M^2 \Sigma$, i.e. effective Planck mass re-scaled $\mu$ and $\Sigma$ parameters. 
\eqref{sigmugam} then yields the following expression for $\hat \Sigma$
\begin{equation} \label{sigmadef}
\hat \Sigma=\frac{\beta_1+\beta_2+\beta_3}{2\beta_1+\beta_2\left(2-\aB\right)}.
\end{equation}
\\

\ni {\bf Implications for GW-constrained Horndeski}:
Using the general results from above and specialising to the case we are focusing on in this paper, i.e. $\aT = 0$ and $\aM = -\aB$, we obtain
\begin{align} \label{betaSol}
\beta_2 &= \aM, &\beta_3 &= \beta_1 + \aM(1+\aM), &\beta_4 &= \tfrac{1}{2}\aM^2.
\end{align}
From this and \eqref{mugamma} it follows that
\begin{align}
\hat\mu - 1 &= \frac{\aM^2}{2\beta_1 + \aM(2+\aM)} = \frac{\aM^2}{2{\cal D}c_s^2}, \nn \\
\gamma - 1 &= - \frac{\aM^2}{\beta_1 + \aM(1+\aM)} = - \frac{\aM^2}{{\cal D}c_s^2 + \tfrac{1}{2}\aM^2},
\label{mugammaSpec}
\end{align}
Since the denominator for both expressions is always positive by virtue of the gradient stability ($c_s^2 > 0$) and ghost freedom (${\cal D} > 0$) conditions, this recovers the conclusion of \cite{Amendola:2017orw}, i.e. that deviations from GR in the model we consider here \eqref{Horndeski_Vsimple} are such that $\gamma \leq 1$ and $\hat \mu \geq 1$, whereas $\hat \Sigma = 1$ identically (this follows from \eqref{sigmadef} and \eqref{betaSol}). 
This is also consistent with the finding of \cite{Pogosian:2016pwr}, that any deviation of $\hat\Sigma - \hat\mu$ from zero (or equivalently here, of $\hat \mu$ from unity) is associated to $\aM \neq 0$ whenever $\aT = 0$.
In addition, note that the only dependence on $c_s^2$ and ${\cal D}$ comes via the combination ${\cal D}c_s^2$, which (unlike the two terms separately) is independent of $\aK$. So this is a direct manifestation of the well-understood fact that $\aK$ drops out in the QSA at leading order \cite{Alonso:2016suf,Lagos:2017hdr}.
Both observables in \eqref{mugammaSpec} fundamentally constrain the (positive or vanishing) ratio of $\aM^2$ and ${\cal D}c_s^2$. We can make this explicit by defining 
\begin{align} \label{aQSdef}
\alpha_{QS} \equiv \aM^2/(2{\cal D} c_s^2)
\end{align}
and accordingly recasting \eqref{mugammaSpec} as
\begin{align}
\hat\mu - 1 &= \alpha_{QS}, &\gamma - 1 &=  -\frac{2\alpha_{QS}}{1 + \alpha_{QS}}.
\label{mugammaSpec2}
\end{align}

Lensing as parametrised by $\Sigma$ is in principle a direct probe of $M$, i.e. of the difference between $\MPl$ and $\MPl^{\rm eff}$. However, a fixed difference between $M$ and $\MPl$ can be absorbed into $\rho\Delta$ and is therefore unobservable at the level considered here\footnote{I thank Emilio Bellini for related discussions.}, which is indeed part of the motivation for using the hatted variables $\hat \mu$ and $\hat \Sigma$. Nevertheless the time-evolution of $M$ (as also quantified by $\aM$) does leave an observable imprint, so lensing over sufficiently long time/distance scales is %in principle 
modified here, whenever an evolving Planck mass is present. From \eqref{mugammaSpec} and \eqref{mugammaSpec2} one can see that clustering observables and measurements of gravitational slip (as parametrised by $\hat\mu$ and $\gamma$) are a more direct probe of the evolution of $\MPl^{\rm eff}$, directly constraining the positive definite (or zero) $\alpha_{QS}$ and hence $\aM$ (and ${\cal D}c_s^2$). %(in terms of ${\cal D}c_s^2$).
\\

\ni {\bf Constraints on quasi-static parameters}:
More quantitatively, we can now use the expressions derived above and the constraints from section \ref{sec-cosmo} (and especially table \ref{tab_am_con}) to place bounds on $\hat\mu$ and $\gamma$. Both $\hat\mu$ and $\gamma$ of course evolve with time in a non-trivial way, mixing the time-dependence of the $\aI$ with other background contributions that enter via $\beta_1$. It will therefore be instructive to first consider the behaviour of $\aQS$ (the single parameter controlling both $\hat\mu$ and $\gamma$) in different limits. We recall the definition of $\aQS$
\eqref{aQSdef} and will work with the $\aM = c_M a$ parametrisation in the context of the GW-constrained Horndeski model \eqref{Horndeski_Vsimple}.\footnote{For a comparison with results for other parametrisations, see appendix \ref{appendix_qsa}.} There we have
\begin{align} %\label{gradient_condition4}
{\cal D}c_s^2 =  - \frac{\dot H}{H^2} \aM + \tfrac{1}{2}\aM^2 - 2\frac{\dot H}{H^2} \frac{(M^2 - 1)}{M^2},
\end{align}
which we can evaluate during a de Sitter phase, a matter- and a radiation-dominated era in analogy to the computation in section \ref{sec-stab}. We find 
\begin{align} 
{\rm dS} &: \quad\quad  {\cal D}c_s^2 = \tfrac{1}{2}\aM^2 \quad \Rightarrow \quad \alpha_{\rm QS} \sim 1  \nn\\
{\rm mat} &: \quad\quad {\cal D}c_s^2 \sim \tfrac{9}{2} \aM \quad \Rightarrow \quad \alpha_{\rm QS} \sim \tfrac{1}{9} \aM  \nn\\
{\rm rad} &: \quad\quad {\cal D}c_s^2 \sim \; 6 \aM \quad \Rightarrow \quad \alpha_{\rm QS} \sim \tfrac{1}{12} \aM.  
\label{aQS_limits_aParam}
\end{align}
To obtain these expressions we have dropped the $\aM^2$ term whenever there is a non-vanishing contribution linear in $\aM$ (due to the smallness of $\aM$ enforced by the constraints of section \ref{sec-cosmo}) and have also used that here $\aM \sim (M^2-1)/M^2$, which we verify in appendix \ref{appendix_qsa}. $\aQS$ and deviations from GR are therefore strongest at very late times, whereas they are strongly suppressed at early times (recall that $\aM$ monotonically increases in time here).

For current experiments at low redshift we of course do not precisely need any of the above limits, but an analogous expression for the value of $\aQS$ today. Using the background equations \eqref{back} as well as the approximate relations $\rho_{{\rm DE},0} \sim 0.7 \rho_{{\rm tot},0}$ and $\rho_{{\rm mat},0} \sim 0.3 \rho_{{\rm tot}, 0}$, we find that $({\dot H}/H^2)|_0 \sim -1/2$, where a $0$ subscript denotes the value of the given quantity today. We therefore find
\begin{align} %\label{gradient_condition4}
\left.{\cal D}c_s^2\right|_0 \sim \tfrac{3}{2} \alpha_{M,0} \quad \Rightarrow \quad \alpha_{{\rm QS}, 0} \sim \tfrac{1}{3} \alpha_{M,0}.
\end{align}
Given the monotonically increasing form of $\aM$ (in the parametrisation considered) and using the constraints in table \ref{tab_am_con}, 
%$\alpha_{{\rm QS},0}$ (and, given its monotonically increasing nature, 
$\alpha_{\rm QS}$ until today satisfies
\begin{align} \label{aQScon}
\alpha_{\rm QS} \lesssim 0.02
\end{align}
at the $2\sigma$ confidence level, so this is the constraint most relevant for current experiments.\footnote{From \eqref{aQS_limits_aParam} one can see, however, that $\aQS$ will continue to grow in this parametrisation in the future -- see appendix \ref{appendix_qsa} for an example, where this growth is much reduced.}
Note that this constraint (just as the constraint for the $c_i$) is somewhat dependent on the $\alpha$-parametrisation chosen (for the $\aM = c_M \Omega_{\rm DE}$ parametrisation it is $\alpha_{\rm QS} \lesssim 0.09$ -- see appendix \ref{appendix_qsa} for details). Once a particular bound is established, however, it can be straightforwardly mapped into a constraint on $\hat\mu$ and $\gamma$. From \eqref{aQScon} we find
\begin{align}
\hat \mu - 1 &\lesssim 0.02, &\gamma -1 &\gtrsim - 0.04,
\end{align}
where we recall that stability bounds also enforce $\gamma \leq 1$ and $\hat \mu \geq 1$.
In other words, the constraint on $\alpha_{\rm QS}$ \eqref{aQScon} implies that deviations from GR as measured by $\hat \mu$ are expected to be at most at the $2\%$ level, while for $\gamma$ they are at most at the $4\%$ level. From this it is also clear that the smallness of $\alpha_{\rm QS}$ implies that we have
\begin{align}
\frac{\gamma -1}{\hat \mu -1} \sim -2
\end{align}
with corrections at the level of $0.04$, i.e. at the $2\%$ level as for $\hat\mu$. The key conclusion of this section is therefore that tight bounds on $\aM$ in the context of the reduced Horndeski theory \eqref{Horndeski_Vsimple} can easily be re-cast as bounds on the quasi-static $\hat\mu$ and $\gamma$ parameters (recall that $\hat\Sigma$ is identically unity for such theories). Given the constraints from section \ref{sec-cosmo} and using $\aM = c_M a$, we therefore expect such quasi-static deviations to at most be at the level of a few percent.
\\

\section{Conclusions} \label{sec-conclusions}
In this paper we explored what constraints can be placed on dark energy in light of recent GW bounds.
More specifically, we recapped the nature of the theoretical constraints derived via measurements of the speed of GWs \cite{Baker:2017hug,Creminelli:2017sry,Sakstein:2017xjx,Ezquiaga:2017ekz}  and from requiring the absence of GW-induced (gradient and ghost) dark energy instabilities \cite{Creminelli:2019kjy}, subsequently using these to draw the following key conclusions:
\begin{itemize}
\item For general Horndeski scalar-tensor theories, observable linear cosmological perturbations are controlled by three free functions ($\aB,\aM,\aT$, with a fourth free function not constrainable at the linear level). The GW priors discussed here can be used to eliminate two combinations of these functions, leaving $\aM = -\aB$ as the only relevant freedom.\footnote{$\aT$ constraint-related caveats to this argument are discussed in section \ref{sec-GWcon}.} 
\item For the surviving theories, observational constraints are largely driven by CMB data (with the other LSS probes considered here only having minimal effect), namely by the ISW effect. Constraints on the residual free $\aM = -\aB$ parameter alone are improved by an order of magnitude, when compared with the case without GW priors. These significantly tightened bounds are already comparable to the constraining power previously only expected from next generation LSS experiments \cite{Alonso:2016suf}.
Figures \ref{fig-comCon} and \ref{fig-1D} and table \ref{tab_am_con} summarise constraints.
\item Apart from CMB measurements, the other key driver of constraints on the surviving theories considered are gradient instabilities on cosmological backgrounds. These rule out negative $\aM$ cosmologies (while the ISW effect rules out $\aM \gtrsim \Oo$). We also discussed the parametrisation-dependent presence of instabilities in the radiation-dominated era, which we hope will help to select increasingly physically informed parametrisations for well-motivated theories going forward.
\item In close analogy to \cite{Noller:2019chl} we pointed out that a natural way of interpreting (and satisfying) the GW priors is as a constraint on the strong coupling scale $\Lambda$ suppressing higher-derivative interactions in Horndeski theories. We highlighted that $\Lambda$ can be raised by one to three orders of magnitude (depending on other priors) to satisfy GW priors, while still yielding a functional Vainshtein mechanism on solar system scales. 
\item Finally, we explored the quasi-static regime (relevant for most observable modes) for the surviving theory space, discussing the form of the quasi-static $\mu,\gamma,\Sigma$ functions and placing observational bounds on them -- see section \ref{sec-qsa} for details.
\end{itemize}   
Recent years have seen phenomenal progress in testing and constraining the presence (and potential nature) of additional light degrees of freedom in cosmology, which are a generic consequence of deviations from GR. These constraints have been driven by a fruitful complementarity of bounds from the CMB, large scale structure and from more local measurements (such as solar system constraints), but have been especially impacted by the advent of GW astronomy. 
What we have shown here is just how significant an improvement on dark energy constraints these new insights from GWs produce. 

Looking ahead in the spirit of the constraints presented here, 
combining theoretical constraints and insights with upcoming new data is a particularly promising avenue going forward. 
Examples illustrating the potential of such an approach include the derivation of joint constraints from theoretical priors together with current experimental bounds, e.g. along the lines explored in \cite{Melville:2019wyy,radstab} for positivity and radiative stability priors, as well as careful (re-)analyses of current GW constraints using EFT techniques, explicitly taking into account the associated energy scales and frequencies, e.g. as in \cite{deRham:2018red}.
In addition, GW `standard siren' tests will become a powerful probe of the conformal $G_4(\phi)R$ coupling in \eqref{Horndeski_Vsimple}, which is the main driver of the surviving Horndeski-related deviations from GR discussed and constrained throughout this paper -- for related discussions see e.g. \cite{Saltas:2014dha,Lombriser:2015sxa,Amendola:2017ovw,Belgacem:2017ihm,Belgacem:2018lbp,Lagos:2019kds,Wolf:2019hun,Belgacem:2019pkk,Dalang:2019fma,Dalang:2019rke,DAgostino:2019hvh,Garoffolo:2019mna}.
Ultimately a holistic analysis, combining constraints from several regimes and using a range of techniques stretching from observational astrophysics to particle theory, will be essential in acquiring the best possible understanding of the dynamics underlying the accelerated expansion of the late universe. We hope the present paper will be a stepping stone on that path.

\section*{Acknowledgments}
\vspace{-0.1in}
\noindent I thank Emilio Bellini, Macarena Lagos, Luca Santoni, Enrico Trincherini, Leonardo Trombetta and Filippo Vernizzi for useful discussions. I would also like to acknowledge support from Dr. Max R\"ossler, the Walter Haefner Foundation and the ETH Zurich Foundation.   
In deriving the results of this paper, I have used: CLASS \cite{Blas:2011rf},  corner \cite{corner}, hi\_class \cite{Zumalacarregui:2016pph}, MontePyton \cite{Audren:2012wb,Brinckmann:2018cvx}, xAct \cite{xAct} and xIST \cite{xIST}.
\\

%%%%%%%%
\appendix
%\section*{Appendix}
%\appendix

\section{Regulating early-time gradient instabilities} \label{appendix_grad}
In section \ref{sec-stab} we showed that a gradient instability generically arises in the radiation-dominated era, for the $\aM = c_m \Omega_{\rm DE}$ parametrisation in the reduced Horndeski model \eqref{Horndeski_Vsimple} (i.e. in Horndeski models consistent with the GW prior).
Is such an instability catastrophic or can it be kept under control? The answer depends on details of the parametrisation and essentially amounts to whether the instability can be sufficiently suppressed. To this end $\aK$, the kineticity parameter we have largely set aside so far, becomes important. In the restricted Horndeski model \eqref{Horndeski_Vsimple} we are focusing on, this satisfies
\begin{align} \label{ak_exp}
M^2 \aK &= 2 \frac{H_0^2}{H^2} X \left( G_{2,X} + 2 X G_{2,XX} \right).
\end{align}
For a simple canonical choice of $G_2 = X$ and assuming $M^2$ is $\Oo$, this means $H^2 \aK \sim H_0^2 X$. In computing parameter constraints, $\aK$ is frequently parametrised with an additional constant parameter in comparison to the other $\aI$ \cite{Zumalacarregui:2016pph}, so for the $\aI = c_i \Omega_{\rm DE}$ parametrisation this means $\aK = d_k + c_k \Omega_{\rm DE}$ (and analogously for other parametrisations). Adding a non-zero $d_k$ is somewhat motivated by the intuition that, while interactions between a proposed DE scalar and other degrees of freedom should switch off at early times (hence e.g. $\aM \to 0$ at early times), one might expect the DE scalar to decouple and keep its canonical kinetic term (which negligibly contributes to the universe's energy density and yields a small $\aK$, whose details depend on the evolution of $X$). In any case, if there is no gradient instability, adding a small $d_k$ typically suppresses $c_s$ (and associated oscillations) at early times, conveniently reducing the time to compute constraints (for other parameters) without changing their values.\footnote{One way to set $d_k$ in {hi\_{}class} is via the so-called {\it kineticity\_{}safe} parameter.} 
This is e.g. the case for the $\aM = c_m a$ constraints for \eqref{Horndeski_Vsimple} shown in section \ref{sec-cosmo} above, which are robust to setting $d_k = 0$ or different fiducial choices of $c_k$. 

When there is a gradient instability in the radiation era, as in the $\aM = c_M \Omega_{\rm DE}$ parametrisation for \eqref{Horndeski_Vsimple}, it is important to recall \eqref{sound_speed}. If $\aK$ is `large' at early times, this drives $c_s^2$ very close to zero (while still negative) during the radiation era, strongly suppressing any developing instabilities. So a `large' $c_k$ or $d_k$ can sufficiently suppress developing instabilities until they are cured at lower redshifts. To give some specific examples and quantify the meaning of `large', running hi\_class with $d_k = 0, c_k = 0.1$, an instability develops for $c_m \gtrsim 0.5$, leading to a catastrophic growth of power on small scales (i.e. large $\ell$). 
With $d_k = 0, c_k = 10^{-2}$ catastrophic instabilities already develop for $c_M \gtrsim 0.03$, while for $d_k \gtrsim 10^{-3}$ all observationally relevant choices of $c_M$ (as well as all above fiducial choices for $c_k$) are essentially unaffected by the early-time instability. 
While the precise details of these bounds on $c_k$ and $d_k$ may have some sensitivity on precision settings of the code, the overall qualitative picture is therefore clear: A sufficiently large $\aK$ (sourced by a non-zero $c_k$ and/or $d_k$) can suppress radiation-era gradient instabilities to the point where they can be neglected.
Constraints for $\aM = c_M \Omega_{\rm DE}$ parametrisation shown in section \ref{sec-cosmo} are accordingly obtained setting $c_k = 0.1$ and $d_k = 10^{-2}$. As a final note, it is important not to over-interpret the above `solution' for dealing with early universe gradient instabilities. It should be seen as a `fix' for an artefact of a specific parametrisation, i.e. a way to suppress the otherwise sick linear dynamics of the scalar at early times (making it behave similarly to a dust component \cite{Zumalacarregui:2016pph}) in this parametrisation. So while e.g. sending $c_s^2$ close to zero is generically associated with strong coupling problems when considering a full non-linear theory, here one shouldn't derive any such fundamental conclusions from artefacts of (the implementation of) parametrisation. As discussed in the main text, increasingly realistic and theoretically informed parametrisations (which will hopefully arise in the near future) should not require such `hacks', so this should be a transient implementation detail in Einstein-Boltzmann codes. 
\\

\section{Additional stability and quasi-static constraints details} \label{appendix_qsa}

\begin{figure}[t]
\begin{flushright}
\includegraphics[width=.99\linewidth]{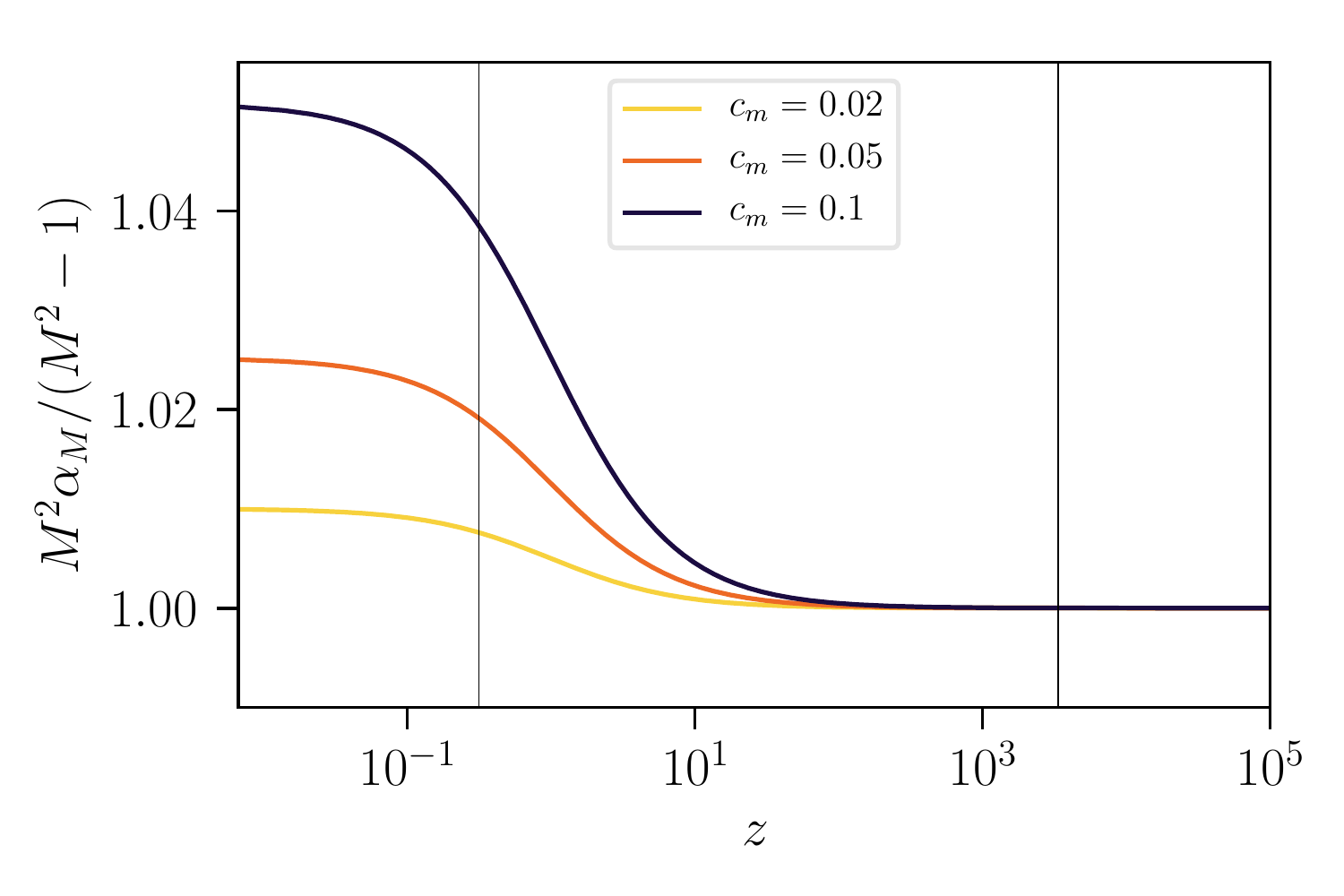} \\
\includegraphics[width=.94\linewidth]{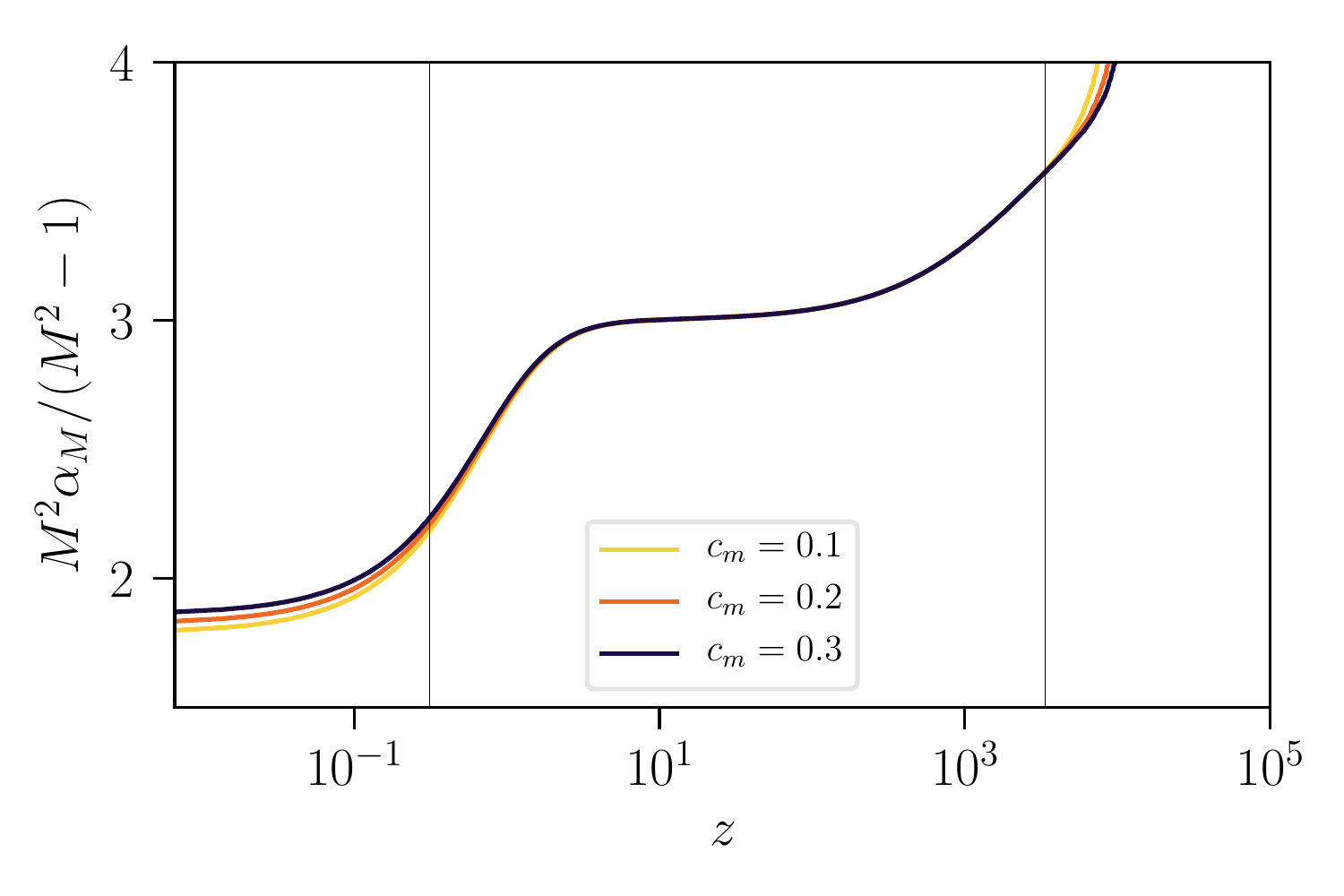}
\end{flushright}
\caption{
Plots of the ratio of $\aM$ and $(M^2-1)/M^2$ vs. redshift $z$ -- cf. the terms in \eqref{app_gradient_condition_4} and \eqref{app_gradient_condition_5} -- using the $\aI = c_i \cdot a$ ({\bf upper plot}) and $\aI = c_i \cdot \Omega_{\rm DE}$ ({\bf lower plot}) parametrisations. The ratio of these quantities during different eras is used to estimate quasi-static observables in section \ref{sec-qsa}. Values for $c_i$ are chosen to span the observationally allowed range of values -- see table \ref{tab_am_con}. Faded vertical lines denote the redshifts of matter-dark energy and matter-radiation equality.
\label{fig-GradRatio}}
\end{figure}

Let us recall the expression for ${\cal D}c_s^2$ as relevant for gradient stability criteria for the two different main parametrisations considered throughout this paper and in the context of the reduced Horndeski theory \eqref{Horndeski_Vsimple}. For $\aM = c_M a$ we have
\begin{align} \label{app_gradient_condition_4}
{\cal D}c_s^2 =  - \frac{\dot H}{H^2} \aM + \tfrac{1}{2}\aM^2 - 2\frac{\dot H}{H^2} \frac{(M^2 - 1)}{M^2} \geq 0,
\end{align}
whereas for $\aM = c_M \Omega_{\rm DE}$ we find
\begin{align} \label{app_gradient_condition_5}
{\cal D}c_s^2 = \aM \left(1 + \frac{\dot H}{H^2}\right) + \tfrac{1}{2}\aM^2 - 2\frac{\dot H}{H^2} \frac{(M^2 - 1)}{M^2} \geq 0.
\end{align}
Given the observational constraints of section \ref{sec-cosmo}, we can drop the $\aM^2$ term in favour of the linear $\aM$ term in both cases (except in the de Sitter limit of the first expression, where the contribution linear in $\aM$ vanishes). A question relevant for the estimates made in section \ref{sec-qsa} now is the relative size of the $\aM$- and $M^2$-dependent terms. In figure \ref{fig-GradRatio} we plot the ratio between $\aM$ and $(M^2-1)/M^2$ for both parametrisations and a number of choices for $c_M$. We find that $\aM \sim (M^2-1)/M^2$ at all times for the $\aM = c_M a$ parametrisation, whereas $\aM M^2/(M^2-1)$ is approximately 2 today and transitions to $\sim 3$ during matter-domination for the $\aM = c_M \Omega_{\rm DE}$ parametrisation (we do not track radiation-domination in detail in this case, since it is plagued by instabilities associated with $\aM M^2/(M^2-1) \gtrsim 4$).

In section \ref{sec-qsa} we used the $\aM = c_M a$ parametrisation to place constraints on the quasi-static parameters $\hat\mu$ and $\gamma$ for \eqref{Horndeski_Vsimple}. To get a feel for the parametrisation-dependence of these results, we here repeat the same exercise for the $\aM = c_M \Omega_{\rm DE}$ parametrisation. ${\cal D}c_s^2$ is now given by \eqref{app_gradient_condition_5}. Evaluating this for different regimes as in section \ref{sec-qsa}, we find the following expressions for $\aQS$ 
\begin{align} 
{\rm dS} &: \quad\quad {\cal D}c_s^2 \sim \aM \quad \;\; \Rightarrow \quad \alpha_{\rm QS} \sim \tfrac{1}{2} \aM  \nn\\
{\rm mat} &: \quad\quad {\cal D}c_s^2 \sim \tfrac{1}{2}\aM \quad \Rightarrow \quad \alpha_{\rm QS}  \sim \aM,
\end{align}
where we do not discuss the radiation-dominated era again, due to the associated instabilities. 
We have dropped the $\aM^2$ term in favour of the (non-vanishing) contribution linear in $\aM$  (due to the smallness of $\aM$ enforced by the constraints of section \ref{sec-cosmo}, as before) and have also used that %$\aM \sim 2 (M^2-1)/M^2$ and 
$\aM \sim 3 (M^2-1)/M^2$ during %dark energy-domination and 
matter-domination in this parametrisation, as discussed above.

With an eye on current experiments we again also compute the corresponding expression for today (as detailed in section \ref{sec-qsa}), obtaining
\begin{align} %\label{gradient_condition4}
\left.{\cal D}c_s^2\right|_0 \sim \alpha_{M,0} \quad \Rightarrow \quad \alpha_{{\rm QS}, 0} \sim \tfrac{1}{2} \alpha_{M,0}.
\end{align}
Using the constraints in table \ref{tab_am_con} in the same way as in the main text, in this parametrisation $\alpha_{\rm QS}$ until today therefore satisfies 
\begin{align} \label{app_aQScon}
\alpha_{\rm QS} \lesssim 0.09
\end{align}
at the $2\sigma$ confidence level, where we emphasise the extra factor of $\Omega_{DE,0}$, which is present in this parameterisation nd which needs to be taken into account when translating bounds from table \ref{tab_am_con} into $\aM$ and ultimately $\aQS$. While $\aQS$ will continue to grow with $\Omega_{\rm DE}$ here, the expression for $\aQS$ in the de Sitter limit is identical to that for today, so there will be no additional formal enhancement as time progresses, unlike for the case discussed in the main text. The constraint \eqref{app_aQScon} now translates into 
\begin{align}
\hat \mu - 1 &\lesssim 0.09, &\gamma -1 &\gtrsim - 0.17,
\end{align}
amounting to deviations from GR at up to the $9\%$ and $17\%$ percent level for $\hat\mu$ and $\gamma$, respectively.

\bibliographystyle{utphys}
\bibliography{noG3ct}
\end{document}